\documentclass[12pt]{article}
\usepackage[a4paper,hmargin=30mm,vmargin={30mm,15mm},
footskip=15mm,footnotesep=9mm,nohead,dvips]{geometry}
\usepackage{amsmath}
\usepackage{amsfonts}
\usepackage{amssymb}
\usepackage{amsthm}

\renewcommand{\leq}{\leqslant}
\renewcommand{\geq}{\geqslant}
\renewcommand{\vec}{\boldsymbol}
\newcommand{\rmd}{\mathrm{d}}
\newcommand{\Etw}{\mathbb{E}^{2}}
\newcommand{\Eth}{\mathbb{E}^{3}}
\newcommand{\eref}[1]{(\ref{#1})}

\newtheorem{proposition}{Proposition}[section]
\newtheorem{theorem}{Theorem}[section]
\newtheorem{definition}{Definition}[section]

\numberwithin{equation}{section}

\begin{document}

\begin{center}
  {\Large Invariant classification of orthogonally separable Hamiltonian systems
    in Euclidean space} \\[2ex plus 0.5ex minus 0.5ex]
  {\large Joshua T.\ Horwood\footnote{Department of Applied Mathematics and
    Theoretical Physics, University of Cambridge, Cambridge, United Kingdom
    CB3~0WA, email:\ j.horwood@damtp.cam.ac.uk},
    Raymond G.\ McLenaghan\footnote{Department of Applied Mathematics,
    University of Waterloo, Waterloo, Ontario, Canada N2L~3G1, email:\
    rgmclena@uwaterloo.ca},
    Roman G.\ Smirnov\footnote{Department of Mathematics and Statistics,
    Dalhousie University, Halifax, Nova Scotia, Canada B3H~3J5, email:\
    smirnov@mathstat.dal.ca}} \\[2ex plus 0.5ex minus 0.5ex]
  {\large November 15, 2004}
\end{center}

\begin{quote}
  {\small \textbf{Abstract.} The problem of the invariant classification of the
  orthogonal coordinate webs defined in Euclidean space is solved within the
  framework of Felix Klein's Erlangen Program. The results are applied to the
  problem of integrability of the Calogero-Moser model.}
\end{quote}


\section{Introduction} \label{sec:intro}

In his famous \textit{Erlangen Program} \cite{Kl72}, Felix Klein introduced a
unified point of view according to which many different branches of geometry
could be integrated into a single system. As is well known, this standpoint
stipulates that the main goal of any branch of geometry can be formulated as
follows:

\begin{quote}
  \textit{``Given a manifold and a group of transformations of the manifold,
  to study the manifold configurations with respect to those features that are
  not altered by the transformations of the group.''} (\cite{Kl93}, p~67)
\end{quote}

The term ``manifold of $n$ dimensions'' in this setting describes a set of $n$
variables that independently take on the real values from $-\infty$ to
$\infty$ (\cite{Kl73}, p~116).

Motivated by this idea, one can assert that Euclidean geometry of $\Eth$
(Euclidean space) can be completely characterized by the invariants of the
Euclidean group of transformations. As is well-known, this Lie group of
(orientation-preserving) isometries, denoted here by $I(\Eth)$, is a
semi-direct product of the corresponding groups of rotations and translations.

An important aspect of Euclidean geometry is the theory of orthogonal
coordinate webs that originated in works of a number of eminent mathematicians
of the past including St\"{a}ckel~\cite{St91}, B\^{o}cher~\cite{Bo94},
Darboux~\cite{D10} and Eisenhart~\cite{E34} within the framework of the theory
of separation of variables. Its modern developments can be found in the review
by Benenti~\cite{Be04} and the relevant references therein. In particular, it
has been shown that there exist exactly eleven orthogonal coordinate webs which
afford separation of variables for the Schr\"{o}dinger and Hamilton-Jacobi
equations defined in $\Eth$. These coordinate webs are confocal quadrics
determined by the Killing tensors of valence two having orthogonally integrable
(normal) eigenvectors and distinct eigenvalues. Eisenhart's results in $\Eth$
were extended by Olevsky~\cite{Ole50} to three-dimensional spaces of non-zero
constant curvature, while Kalnins, Miller and others generalized them to spaces
of higher dimensions (see Kalnins~\cite{Ka86} and the references therein). 
 
This work is a natural continuation of the project initiated in \cite{MST02}
(see also \cite{MST04} and \cite{SY04}) where isometry group invariants and
covariants of valence-two Killing tensors are derived and used to classify
orthogonal coordinate webs of the Euclidean and Minkowski planes. Accordingly,
we approach the problem of classification of the eleven orthogonal webs in
$\Eth$ from the viewpoint of the invariant theory of the isometry group
$I(\Eth)$. Recall that the standard approach to the study of Killing tensors
defined in pseudo-Riemannian manifolds of constant curvature rests on the fact
that they can be expressed in this case as sums of symmetrized tensor products
of Killing vectors. In contrast to the conventional view, we consider the
Killing tensors of valence two defined in $\Eth$ to be algebraic objects or
elements of the corresponding vector space $\mathcal{K}^{2}(\Eth)$ and define
the action of $I(\Eth)$ in this vector space to derive
\textit{$I(\Eth)$-invariants} of the valence-two Killing tensors.

\begin{quote}
  \textit{In line with the postulates of the Erlangen Program, we completely
  solve the problem of classification of the eleven orthogonal webs in
  $\Eth$ by employing the $I(\Eth)$-invariants of the vector space
  $\mathcal{K}^{2}(\Eth)$ and its subspaces.}
\end{quote}
 
Our solution is based on the result of theorem~\ref{theorem:inv} which
describes the space of all isometry group invariants of the vector space of
Killing tensors of valence two defined in $\Eth$ combined with a careful study
of the corresponding vector space of Killing vectors (the Lie algebra of the
isometry group of $\Eth$).

It must be emphasized that the problem of the invariant classification of the
orthogonal coordinate webs in $\Eth$ is significantly more complicated than the
corresponding problems in two-dimensional pseudo-Riemannian spaces of constant
curvature. Apart from the obvious difficulties in dealing with a vector space
of a much higher dimension, one has to solve the problem of the normality of
eigenvectors of valence-two Killing tensors. More specifically, the eleven
orthogonal coordinate webs in $\Eth$ are generated by Killing tensors of
valence two with normal eigenvectors. On the other hand, unlike the situation
in two-dimensional spaces, in $\Eth$ not every Killing tensor of valence two
with distinct eigenvalues has normal eigenvectors. Moreover, the normality
condition is equivalent to a system of non-linear partial differential
equations (PDEs), which makes it nearly impossible to verify directly. The
problem of finding necessary and sufficient \textit{intrinsic} conditions for
the eigenvectors of a tensor field of valence two with pointwise distinct
eigenvalues to be orthogonally integrable has a long history. It can be traced
back to Schouten~\cite{Sc24}, where such conditions depending on the
eigenvectors were derived. Tonolo~\cite{To49} subsequently determined a set of
eigenvector-independent necessary and sufficient conditions for the Ricci
tensor defined in a three-dimensional space to have normal eigenvectors. This
criterion was shown by Schouten to be applicable to arbitrary (Ricci or not)
valence-two tensor fields defined in an arbitrary pseudo-Riemannian manifold.
Later, Nijenhuis~\cite{N51} derived an equivalent formulation of the criterion
introduced originally by Tonolo in terms of the components of the Nijenhuis
tensor of the tensor field in question. In view of their respective
contributions, we refer to these remarkable formulae throughout this paper as
the \textit{Tonolo-Schouten-Nijenhuis (TSN) conditions} and employ them to
verify the normality of the eigenvectors of valence-two Killing tensors defined
in $\Eth$. In addition, we determine in each case the coordinate transformation
from the given Cartesian coordinates to the corresponding coordinate system
determined by the orthogonal web.

As an illustration of the power of the new theory, we use it to obtain a
concise solution to the problem of integrability of the Calogero-Moser
Hamiltonian system defined in $\Eth$ via orthogonal separation of variables in
the associated Hamilton-Jacobi equation.
 
The paper is organized as follows. In section~\ref{sec:ITKT}, we give an
overview of the invariant theory for vector spaces of Killing tensors defined
on pseudo-Riemannian manifolds of constant curvature. This theory is then
specialized in section~\ref{sec:ITK2E3} to vector spaces of valence-two Killing
tensors in Euclidean space. In section~\ref{sec:hj}, we discuss Hamilton-Jacobi
theory in the context of separation of variables and use it to derive canonical
forms for the orthogonal coordinate webs of $\Eth$. The fundamental invariants
are derived in section~\ref{sec:inv} and are used in section~\ref{sec:class}
to classify the coordinate webs. Methods for transforming a given Killing
tensor to canonical form are treated in section~\ref{sec:cform}. In
section~\ref{sec:alg}, we summarize the steps in our algorithm and apply it
in section~\ref{sec:cm} to determine separable coordinates for the
Calogero-Moser system in $\Eth$. Finally, we draw conclusions in
section~\ref{sec:conc} and indicate future research directions.

The reader will no doubt realize that our classification of the coordinate
webs and our algorithm for determining separable coordinates for natural
Hamiltonians in $\Eth$ is highly computational. Nevertheless, all computations
are \textit{purely algebraic} in nature, and thus are straightforward to
implement in a computer algebra system. To complement the paper, we have
written a Maple package, called the \texttt{KillingTensor} package, which
performs all the steps in our algorithm. The package is available through the
Maple Application Centre at \mbox{http://www.mapleapps.com}.


\section{Invariant theory of Killing tensors} \label{sec:ITKT}

In the past decade, the classical invariant theory of homogeneous polynomials
has become an active field of research once again (see Olver~\cite{Olv99} and
the references therein). The theory emerged in the nineteenth century as the
intrinsic study of vector spaces of homogeneous polynomials under the action
of the general linear group. Two of the authors (RGM, RGS) and Dennis The have
incorporated the basic ideas of classical invariant theory into the study of
Killing tensors defined in pseudo-Riemannian spaces of constant curvature under
the action of the isometry group \cite{MST01,MST02,MST03a,MST03b,MST04}. This
synergy of the two theories grew out of the observation that Killing tensors of
the same valence defined in a space of constant curvature constitute a vector
space or, more precisely, a representation space of the isometry group of the
underlying space. Putting this observation in proper perspective allows one to
extend the basic ideas of classical invariant theory to the study of Killing
tensors. Indeed, let $(M, \vec{g})$ be an $n$-dimensional pseudo-Riemannian
manifold of constant curvature with metric tensor $\vec{g}$.

\begin{definition}
  A \textbf{Killing tensor $\vec{K}$ of valence $\boldsymbol{p}$} defined in
  $(M, \vec{g})$ is a symmetric $(p,0)$ tensor satisfying the Killing tensor
  equation
  \begin{equation}
    [\vec{K}, \vec{g}] = 0, \label{eq:ITKT:KTeqn}
  \end{equation}
  where $[,]$ denotes the Schouten bracket \cite{Sc40}. When $p=1$, $\vec{K}$
  is said to be a \textbf{Killing vector} (infinitesimal isometry) and the
  equation~\eref{eq:ITKT:KTeqn} reads
  \begin{equation}
    \mathcal{L}_{\vec{K}} \vec{g} = 0, \label{eq:ITKT:KVeqn}
  \end{equation}
  where $\mathcal{L}$ denotes the Lie derivative operator.
\end{definition}

The Schouten bracket is a real bilinear operator, which property together with
\eref{eq:ITKT:KTeqn} implies that the set $\mathcal{K}^{p}(M)$ of all Killing
tensors of valence $p$ defined in $(M, \vec{g})$ is in fact a vector space.
Its dimension $d$ is determined by the \textit{Delong-Takeuchi-Thompson (DTT)
formula} \cite{D82,Ta83,Th86}
\begin{equation}
  d = \dim \mathcal{K}^{p}(M) = \frac{1}{n} \binom{n+p}{p+1} \binom{n+p-1}{p},
    \quad p \geq 1. \label{eq:ITKT:DTT}
\end{equation}
Therefore the general element of $\mathcal{K}^{p}(M)$ is represented by $d$
arbitrary parameters \linebreak[4] $a^{1}, \ldots, a^{d}$, with respect to an
appropriate basis. Alternatively, this fact can be verified by solving the
corresponding Killing tensor equation \eref{eq:ITKT:KTeqn} with respect to a
fixed system of coordinates, in which case the parameters $a^{1}, \ldots,
a^{d}$ appear as constants of integration in the general form of elements of
$\mathcal{K}^{p}(M)$.

Each element $h$ of the isometry group $I(M)$ induces, by the push forward map,
a non-singular linear transformation $\rho(h)$ of $\mathcal{K}^{p}(M)$. By
theorem~3.5 of McLenaghan \textit{et~al\/}\ \cite{MMS04}, the map
\begin{equation}
  \rho : I(M) \rightarrow GL(\mathcal{K}^{p}(M)) \label{eq:ITKT:rho}
\end{equation}
defines a representation of $I(M)$. Indeed, $\rho$ is a group isomorphism. Once
the form of the general element $\vec{K}$ of $\mathcal{K}^{p}(M)$ is available
with respect to some convenient system of coordinates on $M$, the explicit form
of the transformation $\rho(h) \vec{K}$ (written more succinctly as $h \cdot
\vec{K}$) may be written explicitly in terms of the parameters $a^{1}, \ldots,
a^{d}$. We shall be particularly concerned with the smooth real-valued
functions on $\mathcal{K}^{p}(M)$ that are invariant under the group $I(M)$.
The precise definition of such $I(M)$-invariant functions of
$\mathcal{K}^{p}(M)$ is as follows.

\begin{definition}
  Let $(M, \vec{g})$ be a pseudo-Riemannian manifold of constant curvature.
  Let $p \geq 1$ be fixed. A smooth function $F : \mathcal{K}^{p}(M)
  \rightarrow \mathbb{R}$ is said to be an
  \textbf{$\boldsymbol{I(M)}$-invariant of $\boldsymbol{\mathcal{K}^{p}(M)}$}
  iff it satisfies the condition
  \begin{equation}
    F(h \cdot \vec{K}) = F(\vec{K}) \label{eq:ITKT:invdefn}
  \end{equation}
  for $\vec{K} \in \mathcal{K}^{p}(M)$ and for all $h \in I(M)$.
\end{definition}

The main problem of any invariant theory is to describe the whole space of
invariants of a vector space under the action of a group. To achieve this one
has to determine a set of \textit{fundamental invariants} with the property
that any other invariant is an analytic function of the fundamental invariants
(see \cite{Olv99} for more details). The fundamental theorem of invariants of
a regular Lie group action \cite{Olv99} determines the number of fundamental
invariants needed to define the whole of the space of $I(M)$-invariants.

\begin{theorem}
  Let $G$ be a Lie group acting regularly on an $n$-dimensional manifold $M$
  with $s$-dimensional orbits. Then, in a neighbourhood $N$ of each point
  $p \in M$, there exist $n-s$ functionally independent $G$-invariants
  $\Delta_{1}, \ldots, \Delta_{n-s}$. Any other $G$-invariant $\mathcal{I}$
  defined near $p$ can be locally uniquely expressed as an analytic function
  of the fundamental invariants through $\mathcal{I} = F(\Delta_{1}, \ldots,
  \Delta_{n-s})$.\label{theorem:ITKT}
\end{theorem}

In order to determine the form of the invariants of $\mathcal{K}^{p}(M)$, we
use the fact the the invariance of a function under an entire Lie group is
equivalent to the invariance of the function under the infinitesimal
transformations of the group given by the corresponding Lie algebra. The
precise result is given in the following proposition \cite{Olv99}.

\begin{proposition}
  Let $G$ be a connected Lie group of transformations acting regularly on a
  manifold $M$. A smooth real-valued function $F : M \rightarrow \mathbb{R}$ is
  a $G$-invariant iff
  \begin{equation}
    \vec{v}(F) = 0 \label{eq:ITKT:vF}
  \end{equation}
  for all $p \in M$ and for every infinitesimal generator $\vec{v}$ of $G$.
\end{proposition}

In our application, $G$ is the representation $\rho(I(M))$ defined by
\eref{eq:ITKT:rho} and the condition \eref{eq:ITKT:vF} is equivalent to
\begin{equation}
  \vec{U}_{i}(F) = 0, \quad i = 1, \ldots, r, \label{eq:ITKT:PDEs}
\end{equation}
where the $\vec{U}_{i}$ are vector fields which form a basis of the Lie algebra
of the representation and $r = \mathrm{dim}\, I(M) = \frac{1}{2} n (n+1)$. By
theorem~3.5 of \cite{MMS04}, this Lie algebra is isomorphic to the Lie algebra
of $I(M)$. Such a basis may be computed directly as the basis of the tangent
space to $\rho(I(M))$ at the identity if an explicit form of the representation
is available. According to theorem~\ref{theorem:ITKT} of the present paper, the
general solution of the system of first-order PDEs \eref{eq:ITKT:PDEs} is an
analytic function $F$ of a set of fundamental $I(M)$-invariants. The number of
fundamental invariants is $d-s$, where $d$ is given by \eref{eq:ITKT:DTT} and
$s$ is the dimension of the orbits of $\rho(I(M))$ acting regularly in the
space $\mathcal{K}^{p}(M)$.

To determine $s$ and the subspaces of $\mathcal{K}^{p}(M)$ where the isometry
group $I(M)$ acts with orbits of the same dimension, one can use the result of
the following proposition \cite{Olv99}.

\begin{proposition}
  Let a Lie group $G$ act on $M$ and let $p \in M$. The vector space $S|_{p} =
  \mathrm{span}\{ \vec{U}_{i}|_{p} \:|\: \vec{U}_{i} \in \mathfrak{g} \}$
  spanned by all vector fields determined by the infinitesimal generators at
  $p$ coincides with the tangent space to the orbit $\mathcal{O}_{p}$ of $G$
  that passes through $p$, i.e.\ $S|_{p} = T_{p}(\mathcal{O}_{p})$. In
  particular, the dimension of $\mathcal{O}_{p}$ equals the dimension of
  $S|_{p}$.
\end{proposition}

We are now prepared to apply the theory presented thus far to the vector
space $\mathcal{K}^{2}(\Eth)$.


\section{Invariant theory of Killing tensors of valence two in Euclidean space}
  \label{sec:ITK2E3}

We now specialize the general theory of the previous section to the vector
space $\mathcal{K}^{2}(\Eth)$ of valence-two Killing tensors in Euclidean
space $\Eth$. Recall the following well-known result in \cite{OV90} from
invariant theory. 
\begin{theorem}
  The orbits of a compact linear group acting in a real vector space are
  separated by the fundamental (polynomial) invariants.
\end{theorem}
We first note that in our case the group is non-compact and so in order to
distinguish between the orbits of $I(\Eth)$ acting in the vector space
$\mathcal{K}^{2}(\Eth)$ we need to employ a more elaborate analysis than a
mere computation of a set of fundamental invariants. 

It is well-known that in $\Eth$, as in all manifolds of constant curvature, any
Killing tensor is expressible as a sum of symmetrized products of Killing
vectors. The six Killing vectors in $\Eth$ may be written in Cartesian
coordinates $x^{i}$ viz
\begin{equation}
  \vec{X}_{i} = \frac{\partial}{\partial x^{i}}, \qquad
  \vec{R}_{i} = \epsilon^{k}{}_{ji} x^{j} \vec{X}_{k}, \label{eq:ITK2E3:Kvec}
\end{equation}
for $i=1,2,3$, where $\epsilon_{ijk}$ is the Levi-Civita permutation
tensor\footnote{We are using the summation convention throughout and lowering
and raising indices with the Euclidean metric $g_{ij} = \mathrm{diag}(1,1,1)$
and its inverse $g^{ij}$.}. We also note the commutation relations
\begin{equation}
  [\vec{X}_{i}, \vec{X}_{j}] = 0, \quad [\vec{X}_{i}, \vec{R}_{j}] =
    \epsilon^{k}{}_{ij} \vec{X}_{k}, \quad [\vec{R}_{i}, \vec{R}_{j}] =
    \epsilon^{k}{}_{ij} \vec{R}_{k} . \label{eq:ITK2E3:Kvec_comm}
\end{equation}
Thus the general Killing tensor in $\mathcal{K}^{2}(\Eth)$ may be expressed as
\begin{equation}
  \vec{K} = A^{ij} \vec{X}_{i} \odot \vec{X}_{j} + 2 B^{ij} \vec{X}_{i} \odot
    \vec{R}_{j} + C^{ij} \vec{R}_{i} \odot \vec{R}_{j} ,
    \label{eq:ITK2E3:Ktens}
\end{equation}
where the coefficients $A^{ij}$, $B^{ij}$ and $C^{ij}$ are constant and satisfy
the symmetry properties
\begin{equation}
  A^{ij} = A^{(ij)}, \quad C^{ij} = C^{(ij)}. \label{eq:ITK2E3:symmAC}
\end{equation}
It follows from \eref{eq:ITK2E3:Kvec} and \eref{eq:ITK2E3:Ktens} that the
components of the general Killing tensor in $\mathcal{K}^{2}(\Eth)$ with
respect to the natural basis are given by
\begin{equation}
  K^{ij} = A^{ij} + 2 \epsilon^{(i}{}_{\ell k} B^{j)k} x^{\ell}
    + \epsilon^{i}{}_{mk} \epsilon^{j}{}_{n\ell} C^{k\ell} x^{m} x^{n} .
    \label{eq:ITK2E3:Ktens_compts1}
\end{equation}
For future reference, we give explicitly the six independent components of
$K^{ij}$. Noting the symmetries \eref{eq:ITK2E3:symmAC}, it proves convenient
to set (following \cite{BCR00})
\begin{equation}
  A^{ij} = \begin{pmatrix} a_{1} & \alpha_{3} & \alpha_{2} \\ \alpha_{3} &
    a_{2} & \alpha_{1} \\ \alpha_{2} & \alpha_{1} & a_{3} \end{pmatrix}, \quad
  B^{ij} = \begin{pmatrix} b_{11} & b_{12} & b_{13} \\ b_{21} & b_{22} & b_{23}
    \\ b_{31} & b_{32} & b_{33} \end{pmatrix}, \quad
  C^{ij} = \begin{pmatrix} c_{1} & \gamma_{3} & \gamma_{2} \\ \gamma_{3} &
    c_{2} & \gamma_{1} \\ \gamma_{2} & \gamma_{1} & c_{3} \end{pmatrix}
  \label{eq:ITK2E3:ABCmatrices}
\end{equation}
and $x^{i} = (x,y,z)$. From \eref{eq:ITK2E3:Ktens_compts1} we obtain
\begin{align}\begin{split}
  K^{11} &= a_{1} - 2 b_{12} z + 2 b_{13} y + c_{2} z^{2} + c_{3} y^{2}
    - 2 \gamma_{1} y z, \\
  K^{22} &= a_{2} - 2 b_{23} x + 2 b_{21} z + c_{3} x^{2} + c_{1} z^{2}
    - 2 \gamma_{2} z x, \\
  K^{33} &= a_{3} - 2 b_{31} y + 2 b_{32} x + c_{1} y^{2} + c_{2} x^{2}
    - 2 \gamma_{3} x y, \\
  K^{23} &= \alpha_{1} + b_{31} z - b_{21} y + (b_{22} - b_{33}) x
    + (\gamma_{3} z + \gamma_{2} y - \gamma_{1} x) x - c_{1} y z, \\
  K^{31} &= \alpha_{2} + b_{12} x - b_{32} z + (b_{33} - b_{11}) y
    + (\gamma_{1} x + \gamma_{3} z - \gamma_{2} y) y - c_{2} z x, \\
  K^{12} &= \alpha_{3} + b_{23} y - b_{13} x + (b_{11} - b_{22}) z
    + (\gamma_{2} y + \gamma_{1} x - \gamma_{3} z) z - c_{3} x y.
\end{split}\label{eq:ITK2E3:Ktens_compts2}\end{align}
According to the DTT formula \eref{eq:ITKT:DTT}, the dimension of
$\mathcal{K}^{2}(\Eth)$ is twenty which appears to disagree with
\eref{eq:ITK2E3:ABCmatrices} which lists twenty-one parameters. To reconcile
this, we observe from \eref{eq:ITK2E3:Ktens_compts2} that only the differences
of the diagonal coefficients $b_{11}$, $b_{22}$ and $b_{33}$ are involved.
Defining
\begin{equation}
  \beta_{1} = b_{22} - b_{33}, \quad \beta_{2} = b_{33} - b_{11}, \quad
  \beta_{3} = b_{11} - b_{22}, \label{eq:ITK2E3:beta}
\end{equation}
yields the constraint $\beta_{1} + \beta_{2} + \beta_{3} = 0$, thereby showing
that there are twenty independent parameters. In many of the computations
which follow, it turns out to be more convenient to use the three $b_{ii}$
parameters instead of two of the three $\beta_{i}$. With this in mind, we
(commit an abuse of notation and) let $\mathcal{K}^{2}(\Eth)$ be the space
spanned by the twenty-one parameters
\begin{equation}
  a_{1}, a_{2}, a_{3}, \alpha_{1}, \alpha_{2}, \alpha_{3},
  b_{11}, b_{22}, b_{33}, b_{23}, b_{31}, b_{12}, b_{32}, b_{13}, b_{21},
  c_{1}, c_{2}, c_{3}, \gamma_{1}, \gamma_{2}, \gamma_{3} .
    \label{eq:ITK2E3:KTparams}
\end{equation}
We shall also refer to \eref{eq:ITK2E3:KTparams} as the \textit{Killing tensor
parameters}.

We now consider the transformation rules for the Killing tensor parameters.
The transformation from one set of Cartesian coordinates $x^{i}$ to another
set $\tilde{x}^{i}$ is given by
\begin{equation}
  x^{i} = \lambda_{j}{}^{i} \tilde{x}^{j} + \delta^{i},
    \label{eq:ITK2E3:trans_x}
\end{equation}
where $\lambda_{j}{}^{i} \in SO(3)$ and $\delta^{i} \in \mathbb{R}^{3}$. It is
straightforward to show that the Killing vectors \eref{eq:ITK2E3:Kvec}
transform according to
\begin{equation}
  \vec{X}_{i} = \lambda^{j}{}_{i} \vec{\tilde{X}}_{j}, \quad
  \vec{R}_{i} = \lambda^{j}{}_{i} \vec{\tilde{R}}_{j} + \mu^{j}{}_{i}
    \vec{\tilde{X}}_{j}, \label{eq:ITK2E3:trans_Kvec}
\end{equation}
where
\begin{equation}
  \mu^{j}{}_{i} = \epsilon^{k}{}_{\ell i} \lambda^{j}{}_{k} \delta^{\ell} .
    \label{eq:ITK2E3:mu}
\end{equation}
The Killing vector transformation rules \eref{eq:ITK2E3:trans_Kvec} in
conjunction with \eref{eq:ITK2E3:Ktens} lead to
\begin{align}\begin{split}
  \tilde{A}^{ij} &= A^{k\ell} \lambda^{i}{}_{k} \lambda^{j}{}_{\ell}
    + 2 B^{k\ell} \lambda^{(i}{}_{k} \mu^{j)}{}_{\ell}
    + C^{k\ell} \mu^{i}{}_{k} \mu^{j}{}_{\ell}, \\
  \tilde{B}^{ij} &= B^{k\ell} \lambda^{i}{}_{k} \lambda^{j}{}_{\ell}
    + C^{k\ell} \lambda^{j}{}_{\ell} \mu^{i}{}_{k}, \\
  \tilde{C}^{ij} &= C^{k\ell} \lambda^{i}{}_{k} \lambda^{j}{}_{\ell} .
\end{split}\label{eq:ITK2E3:trans_ABC}\end{align}
These equations give the explicit form of the representation of $I(\Eth)$ on
$\mathcal{K}^{2}(\Eth)$ with respect to a Cartesian coordinate system on
$\Eth$.

Equipped with these transformation rules, we can now derive the infinitesimal
generators of $I(\Eth)$ in the representation defined by
\eref{eq:ITK2E3:trans_ABC}. Let $\vec{U}_{m}$, $m=1,2,3$, denote the generators
associated to the Killing vectors $\vec{X}_{m}$. Noting that such Killing
vectors generate translations about the $x^{m}$-axis, we set $\lambda_{j}{}^{i}
= \delta_{j}{}^{i}$ in \eref{eq:ITK2E3:trans_ABC} and differentiate the
resulting equations with respect to $\delta^{m}$ to obtain
$$
  \left. \frac{\partial \tilde{A}^{ij}}{\partial \delta^{m}}
    \right|_{\delta^{i} = 0} = 2 \epsilon^{(i}{}_{mk} B^{j)k}, \quad
  \left. \frac{\partial \tilde{B}^{ij}}{\partial \delta^{m}}
    \right|_{\delta^{i} = 0} = \epsilon^{i}{}_{mk} C^{jk}, \quad
  \left. \frac{\partial \tilde{C}^{ij}}{\partial \delta^{m}}
    \right|_{\delta^{i} = 0} = 0 .
$$
The corresponding differential operators are therefore
\begin{equation}
  \vec{U}_{i} = 2 \epsilon^{(j}{}_{i\ell} B^{k)\ell} \frac{\partial}{\partial
    A^{jk}} + \epsilon^{j}{}_{i\ell} C^{k\ell} \frac{\partial}{\partial
    B^{jk}}, \label{eq:ITK2E3:U}
\end{equation}
for $i=1,2,3$, where the range of summation over the derivative operators is
understood to be over only those parameters listed in \eref{eq:ITK2E3:KTparams}.
Next, let $\vec{V}_{m}$, $m=1,2,3$, denote the generators associated to the
Killing vectors $\vec{R}_{m}$, the generators of rotations about the
$x^{m}$-axis. For an infinitesimal rotation about the $x^{3}$-axis by an
angle $\theta^{3}$, the rotation $\lambda_{j}{}^{i} \in SO(3)$ is given by
$$
  \lambda_{j}{}^{i} = \begin{pmatrix} \cos \theta^{3} & -\sin \theta^{3} & 0
    \\ \sin \theta^{3} & \cos \theta^{3} & 0 \\ 0 & 0 & 1 \end{pmatrix}_{ij}
    \quad\Rightarrow\quad \left. \frac{\rmd \lambda_{j}{}^{i}}{\rmd \theta^{3}}
    \right|_{\theta^{3}=0} = \begin{pmatrix} 0 & -1 & 0 \\ 1 & 0 & 0 \\
    0 & 0 & 0 \end{pmatrix}_{ij} = \epsilon_{3j}{}^{i} .
$$
More generally, for an infinitesimal rotation about the $x^{m}$-axis,
$$
  \left. \frac{\rmd \lambda_{j}{}^{i}}{\rmd \theta^{m}} \right|_{\theta^{m}=0}
    = \epsilon_{mi}{}^{j} \quad\Leftrightarrow\quad
  \left. \frac{\rmd \lambda^{i}{}^{j}}{\rmd \theta^{m}} \right|_{\theta^{m}=0}
    = \epsilon^{i}{}_{jm} .
$$
It thus follows from \eref{eq:ITK2E3:trans_ABC} that
\begin{align}\begin{split}
  \vec{V}_{i} &= (\epsilon^{j}{}_{\ell i} A^{\ell k} + \epsilon^{k}{}_{\ell i}
    A^{j\ell}) \frac{\partial}{\partial A^{jk}} + (\epsilon^{j}{}_{\ell i}
    B^{\ell k} + \epsilon^{k}{}_{\ell i} B^{j\ell}) \frac{\partial}{\partial
    B^{jk}} \\
  &\qquad + (\epsilon^{j}{}_{\ell i} C^{\ell k} + \epsilon^{k}{}_{\ell i}
    C^{j\ell}) \frac{\partial}{\partial C^{jk}},
\end{split}\label{eq:ITK2E3:V}\end{align}
for $i=1,2,3$. As required by the general theory, the generators
\eref{eq:ITK2E3:U} and \eref{eq:ITK2E3:V} satisfy the same commutation
relations as the Killing vectors \eref{eq:ITK2E3:Kvec}, namely
$$
  [\vec{U}_{i}, \vec{U}_{j}] = 0, \quad [\vec{U}_{i}, \vec{V}_{j}] =
    \epsilon^{k}{}_{ij} \vec{U}_{k}, \quad [\vec{V}_{i}, \vec{V}_{j}] =
    \epsilon^{k}{}_{ij} \vec{V}_{k} .
$$
For computational purposes, we shall require explicit expressions for the
generators. It follows from \eref{eq:ITK2E3:U} and \eref{eq:ITK2E3:V} that
\begin{equation}
  \vec{U}_{i} = \sum_{j=1}^{21} \mathcal{G}_{i}{}^{j} \frac{\partial}{\partial
    a^{j}}, \quad \vec{V}_{i} = \sum_{j=1}^{21} \mathcal{G}_{i+3}{}^{j}
    \frac{\partial}{\partial a_{j}}, \label{eq:ITK2E3:gen_explicit}
\end{equation}
for $i=1,2,3$, where $a^{j}$, $j=1,\ldots,21$, are the twenty-one Killing
tensor parameters ordered by \eref{eq:ITK2E3:KTparams} and
\small
$$
  \mathcal{G}_{i}{}^{j} = \left( \begin{array}{cccccccc}
    0 & -2 b_{23} & 2 b_{32} & b_{22} - b_{33} & b_{12} & -b_{13}
      & 0 & -\gamma_{1} \\
    2 b_{13} & 0 & -2 b_{31} & -b_{21} & b_{33} - b_{11} & b_{23}
      & \gamma_{2} & 0 \\
    -2 b_{12} & 2 b_{21} & 0 & b_{31} & -b_{32} & b_{11} - b_{22}
      & -\gamma_{3} & \gamma_{3} \\
    0 & 2 \alpha_{1} & -2 \alpha_{1} & a_{3} - a_{2} & -\alpha_{3} & \alpha_{2}
      & 0 & b_{23} + b_{32} \\
    -2 \alpha_{2} & 0 & 2 \alpha_{2} & \alpha_{3} & a_{1} - a_{3} & -\alpha_{1}
      & -b_{31} - b_{13} & 0 \\
    2 \alpha_{3} & -2 \alpha_{3} & 0 & -\alpha_{2} & \alpha_{1} & a_{2} - a_{1}
      & b_{12} + b_{21} & -b_{12} -b_{21}
  \end{array} \right.
$$
$$
  \begin{array}{ccccccc}
    \gamma_{1} & -c_{3} & \gamma_{3} & 0 & c_{2} & 0 & -\gamma_{2} \\
    -\gamma_{2} & 0 & -c_{1} & \gamma_{1} & -\gamma_{3} & c_{3} & 0 \\
    0 & \gamma_{2} & 0 & -c_{2} & 0 & -\gamma_{1} & c_{1} \\
    -b_{23} - b_{32} & b_{33} - b_{22} & -b_{21} & b_{13} & b_{33} - b_{22}
      & -b_{12} & b_{31} \\
    b_{31} + b_{13} & b_{21} & b_{11} - b_{33} & -b_{32} & b_{12}
      & b_{11} - b_{33} & -b_{23} \\
    0 & -b_{13} & b_{32} & b_{22} - b_{11} & -b_{31} & b_{23} & b_{22} - b_{11}
  \end{array}
$$
$$
  \left. \begin{array}{cccccc}
    0 & 0 & 0 & 0 & 0 & 0 \\
    0 & 0 & 0 & 0 & 0 & 0 \\
    0 & 0 & 0 & 0 & 0 & 0 \\
    0 & 2 \gamma_{1} & -2 \gamma_{1}
      & c_{3} - c_{2} & -\gamma_{3} & \gamma_{2} \\
    -2 \gamma_{2} & 0 & 2 \gamma_{2}
      & \gamma_{3} & c_{1} - c_{3} & -\gamma_{1} \\
    2 \gamma_{3} & -2 \gamma_{3} & 0
      & -\gamma_{2} & \gamma_{1} & c_{2} - c_{1}
  \end{array} \right)_{ij} .
$$
\normalsize Finally, we observe that the coefficient matrix
$\mathcal{G}_{i}{}^{j}$ has rank six almost everywhere, and so, in view of
theorem~\ref{theorem:ITKT}, we expect fifteen fundamental $I(\Eth)$-invariants.
The computation and presentation of these invariants are treated in
section~\ref{sec:inv}.


\section{Hamilton-Jacobi theory and orthogonal coordinate webs} \label{sec:hj}

Consider a Hamiltonian system defined on $(M, \vec{g})$ by a natural
Hamiltonian function of the form
\begin{equation}
  H = \tfrac{1}{2} g^{ij}(\vec{x}) p_{i} p_{j} + V(\vec{x}),
    \quad i, j = 1, \ldots n, \label{eq:hj:natH}
\end{equation}
with respect to the canonical Poisson bi-vector $\vec{P} = \sum_{i=1}^{n}
\frac{\partial}{\partial x^{i}} \wedge \frac{\partial}{\partial p^{i}}$ given
in terms of the position-momenta coordinates $(\vec{x}, \vec{p}) = (x^{i},
p_{i})$, $i=1, \ldots, n$ on the cotangent bundle $T^{\ast}(M)$. As is
well-known, in many cases the Hamiltonian system defined by \eref{eq:hj:natH}
can be integrated by quadratures by finding a complete integral $W$ of the 
corresponding Hamilton-Jacobi equation which is a first-order PDE given by 
\begin{equation}
  \frac{1}{2} g^{ij}(\vec{x}) \frac{\partial W}{\partial x^{i}}
    \frac{\partial W}{\partial x^{j}} + V(\vec{x}) = E, \quad
    p_{j} = \frac{\partial W}{\partial x^{i}}. \label{eq:hj:HJeqn}
\end{equation}
The geometrical meaning of equation~\eref{eq:hj:HJeqn} and its complete
integral $W$ is well-under\-stood (see, for example, \cite{Be03}). Thus, if the
function $F = \frac{1}{2} g^{ij} W_{,i} W_{,j} + V - E = 0$ is regular on the
cotangent bundle $T^{\ast}(M)$, then equation \eref{eq:hj:HJeqn} defines a
hypersurface in $T^{\ast}(M)$. Furthermore, $W$ is a complete integral of
\eref{eq:hj:HJeqn} iff the Lagrangian submanifold $\mathcal{S} \subset
T^{\ast}(M)$ determined by the equations $p_{i} = W_{,i}$ lies on the
hypersurface defined by the Hamilton-Jacobi equation \eref{eq:hj:HJeqn}. 
Solving \eref{eq:hj:HJeqn} is normally based on finding a canonical
transformation to separable coordinates: $(\vec{x}, \vec{p}) \rightarrow
(\vec{u}, \vec{v})$ with respect to which the equation can be solved under the
additive separation ansatz $W(\vec{u}; \vec{c}) = \sum_{i=1}^{n} W_{i}(u^i;
\vec{c})$ and the non-degeneracy condition $\det( \partial^{2} W /\partial
u^{i} \partial c_{j} )_{n \times n} \not= 0$, where $\vec{c} = (c_{1}, \ldots,
c_{n})$ is a constant vector. Orthogonal separation of variables occurs in the
case when the transformations to separable coordinates are
point-transformations and the metric tensor $\vec{g}$ is diagonal with respect
to the coordinates of separation $(\vec{u}, \vec{v})$. A useful criterion for
orthogonal separability is given by Benenti \cite{Be04}.
\begin{theorem}
  The Hamiltonian system defined by \eref{eq:hj:natH} is orthogonally separable
  if and only if there exists a valence-two Killing tensor $\vec{K}$ with (i)
  pointwise simple and real eigenvalues, (ii) orthogonally integrable (normal)
  eigenvectors and (iii) such that 
  \begin{equation}
    \rmd(\vec{K}\, \rmd V) = 0. \label{eq:hj:dKdV}
  \end{equation}\label{theorem:Benenti}
\end{theorem}
A Killing tensor satisfying conditions (i) and (ii) of
theorem~\ref{theorem:Benenti} is called a \textit{characteristic Killing tensor
(CKT)}.

Let us elaborate on conditions~(i) and (ii) in theorem~\ref{theorem:Benenti}.
On two-dimensional Riemannian manifolds of constant curvature these conditions
are trivial, since every eigenvector $\vec{\xi}$ of $\vec{K}$ is normal and
$\vec{K}$ has repeated eigenvalues iff it is a multiple of the metric
$\vec{g}$. In three-dimensions, $\Eth$ in particular, the situation is far more
complicated. Computing eigenvectors of a symmetric $3 \times 3$ tensor is
tedious and becomes virtually intractable if one considers Killing tensors with
arbitrary parameters. Instead, we employ the Tonolo-Schouten-Nijenhuis (TSN)
conditions, as introduced in section~\ref{sec:intro}, which are both necessary
and sufficient for a given symmetric (Killing) tensor field to have integrable
eigenvectors. These conditions read
\begin{subequations}\begin{align}
  & N^{\ell}{}_{[jk} g_{i]\ell} = 0, \label{eq:hj:tsn1} \\
  & N^{\ell}{}_{[jk} K_{i]\ell} = 0, \label{eq:hj:tsn2} \\
  & N^{\ell}{}_{[jk} K_{i]m} K^{m}{}_{\ell} = 0 \label{eq:hj:tsn3}
\end{align}\label{eq:hj:tsn}\end{subequations}
where $N^{i}{}_{jk}$ are the components of the Nijenhuis tensor of $K^{ij}$
given by
\begin{equation}
  N^{i}{}_{jk} = K^{i}{}_{\ell} K^{\ell}{}_{[j,k]} + K^{\ell}{}_{[j}
    K^{i}{}_{k],\ell} . \label{eq:hj:nijenhuis}
\end{equation}
We remark that the TSN conditions \eref{eq:hj:tsn1}--\eref{eq:hj:tsn3} yield
10 quadratic, 35 cubic and 84 quartic equations, respectively, in the Killing
tensor parameters. It is thus straightforward to verify, using
\eref{eq:hj:tsn}, if a given Killing tensor satisfies condition~(ii) of
theorem~\ref{theorem:Benenti}. However, we have been unable to solve the
conditions \eref{eq:hj:tsn} directly to obtain the most general Killing tensor
admitting orthogonally integrable eigenvectors\footnote{Steve Czapor (private
communication) has simplified the situation considerably. Using Gr\"{o}bner
basis theory, he has shown that \eref{eq:hj:tsn1} and \eref{eq:hj:tsn2} imply
\eref{eq:hj:tsn3}, for any Killing tensor $\vec{K} \in \mathcal{K}^{2}
(\Eth)$.}. Condition~(i), namely the distinct eigenvalues condition, like
condition~(ii), can also be verified directly for a given Killing tensor: we
simply compute the discriminant of the characteristic polynomial of $K^{ij}$
and verify that it does not vanish identically.

Although a general solution of the TSN conditions~\eref{eq:hj:tsn} appears
intractable, we may instead employ Eisenhart's method \cite{E34} to derive all
Killing tensors with normal eigenvectors up to equivalence. In particular,
each representative Killing tensor characterizes separability of the
Hamilton-Jacobi equation~\eref{eq:hj:HJeqn} in one of the eleven (orthogonally)
separable coordinate systems in $\Eth$. The Eisenhart method can be described
as follows. Consider the Euclidean metric
$$
  \rmd s^{2} = g_{11} (\rmd u^{1})^{2} + g_{22} (\rmd u^{2})^{2}
    + g_{33} (\rmd u^{3})^{2},
$$
with respect to separable coordinates $u^{i}$. The method yields three
canonical Killing tensors given by
$$
  K_{ij} = \mathrm{diag}(\lambda_{1}\, g_{11}, \lambda_{2}\, g_{22},
    \lambda_{3}\, g_{33}),
$$
where the $\lambda_{i}$ satisfy the linear system of PDEs
\begin{equation}
  \frac{\partial \lambda_{i}}{\partial u^{j}} = (\lambda_{i} - \lambda_{j})
    \frac{\partial}{\partial u^{j}} \ln g_{ii}, \quad
    \text{(no sum)} \label{eq:hj:Eisenhart}
\end{equation}
for $i,j = 1,2,3$ (see \cite{E34}, equations~(1.8)). Trivially, any multiple
of the metric $\vec{g}$ satisfies \eref{eq:hj:Eisenhart}. It is
straightforward to solve \eref{eq:hj:Eisenhart} for each of the eleven
separable coordinate systems in $\Eth$ to obtain the two additional
(non-trivial) canonical Killing tensors which we shall label $\vec{K}_{1}$
and $\vec{K}_{2}$.

We now summarize the results of this calculation. For each of the eleven
separable coordinate systems, we give the corresponding coordinate
transformation, ranges of the separable coordinates, the metric and the
components of the canonical Killing tensors $\vec{K}_{1}$ and $\vec{K}_{2}$.
\renewcommand{\arraystretch}{1.3}
\begin{equation}
  \begin{array}{c} \text{Cartesian:} \\ (x, y, z) \\ \text{(\theequation)}
    \end{array} \quad
  \left\{ \begin{array}{l}
    x = x, \: y = y, \: z = z \\
    -\infty < x,y,z < \infty \\
    \rmd s^{2} = \rmd x^{2} + \rmd y^{2} + \rmd z^{2} \\
    K^{ij}_{1} = \mathrm{diag}(0, 1, 0) \\
    K^{ij}_{2} = \mathrm{diag}(0, 0, 1)
  \end{array} \right. \notag \label{eq:hj:sc1}
\end{equation}\stepcounter{equation}
\begin{equation}
  \begin{array}{c} \text{Circular cylindrical:} \\ (r, \theta, z) \\
    \text{(\theequation)} \end{array} \quad
  \left\{ \begin{array}{l}
    x = r \cos \theta, \: y = r \sin \theta, \: z = z \\
    r \geq 0, \quad 0 \leq \theta < 2 \pi, \quad -\infty < z < \infty \\
    \rmd s^{2} = \rmd r^{2} + r^{2}\, \rmd \theta^{2} + \rmd z^{2} \\
    K^{ij}_{1} = \mathrm{diag}(0, r^{4}, 0) \\
    K^{ij}_{2} = \mathrm{diag}(0, 0, 1)
  \end{array} \right. \notag \label{eq:hj:sc2}
\end{equation}\stepcounter{equation}
\begin{equation}
  \begin{array}{c} \text{Parabolic cylindrical:} \\ (\mu, \nu, z) \\
    \text{(\theequation)} \end{array} \quad
  \left\{ \begin{array}{l}
    x = \frac{1}{2}(\mu^{2}-\nu^{2}), \: y = \mu\nu, \:  z = z \\
    \mu \geq 0, \quad -\infty < \nu < \infty, \quad -\infty < z < \infty \\
    \rmd s^{2} = (\mu^{2} + \nu^{2})(\rmd \mu^{2} + \rmd \nu^{2})
      + \rmd z^{2} \\
    K^{ij}_{1} = \mathrm{diag}(\nu^{2} g_{11}, -\mu^{2} g_{22}, 0) \\
    K^{ij}_{2} = \mathrm{diag}(0, 0, 1)
  \end{array} \right. \notag \label{eq:hj:sc3}
\end{equation}\stepcounter{equation}
\begin{equation}
  \begin{array}{c} \text{Elliptic-hyperbolic:} \\ (\eta, \psi, z) \\
    \text{(\theequation)} \end{array} \quad
  \left\{ \begin{array}{l}
    x = a \cosh \eta \cos \psi, \: y = a \sinh \eta \sin \psi, \: z = z \\
    \eta \geq 0, \quad 0 \leq \psi < 2 \pi, \quad -\infty < z < \infty,
      \quad a > 0 \\
    \rmd s^{2} = a^{2}(\cosh^{2} \eta - \cos^{2} \psi)(\rmd \eta^{2} +
      \rmd \psi^{2}) + \rmd z^{2} \\
    K^{ij}_{1} = \mathrm{diag}(a^{2} \cos^{2} \psi\, g_{11}, a^{2} \cosh^{2}
      \eta\, g_{22}, 0) \\
    K^{ij}_{2} = \mathrm{diag}(0, 0, 1)
  \end{array} \right. \notag \label{eq:hj:sc4}
\end{equation}\stepcounter{equation}
\begin{equation}
  \begin{array}{c} \text{Spherical:} \\ (r, \theta, \phi) \\
    \text{(\theequation)} \end{array} \quad
  \left\{ \begin{array}{l}
    x = r \sin \theta \cos \phi, \: y = r \sin \theta \sin \phi,
      \: z = r \cos \theta \\
    r \geq 0, \quad 0 \leq \theta < \pi, \quad 0 \leq \phi < 2 \pi \\
    \rmd s^{2} = \rmd r^{2} + r^{2}\, \rmd \theta^{2} + r^{2} \sin^{2}
      \theta\, \rmd \phi^{2} \\
    K^{ij}_{1} = \mathrm{diag}(0, r^{4}, r^{4} \sin^{2} \theta) \\
    K^{ij}_{2} = \mathrm{diag}(0, 0, r^{4} \sin^{4} \theta)
  \end{array} \right. \notag \label{eq:hj:sc5}
\end{equation}\stepcounter{equation}
\begin{equation}
  \begin{array}{c} \text{Prolate} \\ \text{spheroidal:} \\ (\eta, \theta, \psi)
    \\ \text{(\theequation)} \end{array} \:\:
  \left\{ \begin{array}{l}
    x = a \sinh \eta \sin \theta \cos \psi, \:
      y = a \sinh \eta \sin \theta \sin \psi, \:
      z = a \cosh \eta \cos \theta \\
    \eta \geq 0, \quad 0 \leq \theta < \pi, \quad 0 \leq \psi < 2 \pi,
      \quad a > 0 \\
    \rmd s^{2} = a^{2}(\sinh^{2} \eta + \sin^{2} \theta)(\rmd \eta^{2} +
      \rmd \theta^{2}) + a^{2} \sinh^{2} \eta \sin^{2} \theta\,
      \rmd \psi^{2} \\
    K^{ij}_{1} = \mathrm{diag}\big( -a^{2} \sin^{2} \theta\, g_{11},
      a^{2} \sinh^{2} \eta\, g_{22}, a^{2}(\sinh^{2} \eta - \sin^{2} \theta)
      g_{33} \big) \\
    K^{ij}_{2} = \mathrm{diag}(0, 0, a^{2} \sinh^{2} \eta \sin^{2} \theta\,
      g_{33} )
  \end{array} \right. \notag \label{eq:hj:sc6}
\end{equation}\stepcounter{equation}
\begin{equation}
  \begin{array}{c} \text{Oblate} \\ \text{spheroidal:} \\ (\eta, \theta, \psi)
    \\ \text{(\theequation)} \end{array} \:\:
  \left\{ \begin{array}{l}
    x = a \cosh \eta \sin \theta \cos \psi, \:
      y = a \cosh \eta \sin \theta \sin \psi, \:
      z = a \sinh \eta \cos \theta \\
    \eta \geq 0, \quad 0 \leq \theta < \pi, \quad 0 \leq \psi < 2 \pi,
      \quad a > 0 \\
    \rmd s^{2} = a^{2}(\cosh^{2} \eta - \sin^{2} \theta)(\rmd \eta^{2} +
      \rmd \theta^{2}) + a^{2} \cosh^{2} \eta \sin^{2} \theta\,
      \rmd \psi^{2} \\
    K^{ij}_{1} = \mathrm{diag}\big( a^{2} \sin^{2} \theta\, g_{11},
      a^{2} \cosh^{2} \eta\, g_{22}, a^{2}(\cosh^{2} \eta + \sin^{2} \theta)
      g_{33} \big) \\
    K^{ij}_{2} = \mathrm{diag}(0, 0, a^{2} \cosh^{2} \eta \sin^{2} \theta\,
      g_{33} )
  \end{array} \right. \notag \label{eq:hj:sc7}
\end{equation}\stepcounter{equation}
\begin{equation}
  \begin{array}{c} \text{Parabolic:} \\ (\mu, \nu, \psi) \\
    \text{(\theequation)} \end{array} \quad
  \left\{ \begin{array}{l}
    x = \mu \nu \cos \psi, \: y = \mu \nu \sin \psi,
      \: z = \frac{1}{2}(\mu^{2}-\nu^{2}) \\
    \mu \geq 0, \quad \nu \geq 0, \quad 0 \leq \psi < 2 \pi \\
    \rmd s^{2} = (\mu^{2} + \nu^{2})(\rmd \mu^{2} + \rmd \nu^{2}) +
      \mu^{2} \nu^{2}\, \rmd \psi^{2} \\
    K^{ij}_{1} = \mathrm{diag}\big( -\nu^{2} g_{11}, \mu^{2} g_{22},
      (\mu^{2} - \nu^{2}) g_{33} \big) \\
    K^{ij}_{2} = \mathrm{diag}(0, 0, \mu^{2}\nu^{2} g_{33})
  \end{array} \right. \notag \label{eq:hj:sc8}
\end{equation}\stepcounter{equation}
\begin{equation}
  \begin{array}{c} \text{Conical:} \\ (r, \theta, \lambda) \\
    \text{(\theequation)} \end{array} \:\:
  \left\{ \begin{array}{l}
    \displaystyle x^{2} = \left(\frac{r\theta\lambda}{bc}\right)^{2} \!\!\!,
      \: y^{2} = \frac{r^{2}(\theta^{2} - b^{2})(b^{2} - \lambda^{2})}
      {b^{2}(c^{2} - b^{2})}, \: z^{2} = \frac{r^{2}(c^{2} - \theta^{2})
      (c^{2} - \lambda^{2})}{b^{2}(c^{2} - b^{2})} \\
    r \geq 0, \quad b^{2} < \theta^{2} < c^{2}, \quad 0 < \lambda^{2}
      < b^{2}, \\
    \displaystyle \rmd s^{2} = \rmd r^{2} + \frac{r^{2}(\theta^{2} -
      \lambda^{2})}{(\theta^{2} - b^{2})(c^{2} - \theta^{2})}\, \rmd \theta^{2}
      + \frac{r^{2}(\theta^{2} - \lambda^{2})}{(b^{2} - \lambda^{2})
      (c^{2} - \lambda^{2})}\, \rmd \lambda^{2} \\
    K^{ij}_{1} = \mathrm{diag}(0, r^{2}\lambda^{2} g_{22},
      r^{2} \theta^{2} g_{33}) \\
    K^{ij}_{2} = \mathrm{diag}(0, r^{2} g_{22}, r^{2} g_{33})
  \end{array} \right. \notag \label{eq:hj:sc9}
\end{equation}\stepcounter{equation}
\begin{equation}
  \begin{array}{c} \text{Paraboloidal:} \\ (\mu, \nu, \lambda) \\
    \text{(\theequation)} \end{array} \quad
  \left\{ \begin{array}{l}
    \displaystyle x^{2} = \frac{4(\mu-b)(b-\nu)(b-\lambda)}{b-c}, \:
      y^{2} = \frac{4(\mu-c)(c-\nu)(\lambda-c)}{b-c}, \\
      \qquad z = \mu + \nu + \lambda - b - c \\
    0 < \nu < c < \lambda < b < \mu < \infty \\
    \displaystyle \rmd s^{2} = \frac{(\mu-\nu)(\mu-\lambda)}{(\mu-b)(\mu-c)}\,
      \rmd \mu^{2} + \frac{(\mu-\nu)(\lambda-\nu)}{(b-\nu)(c-\nu)}\,
      \rmd \nu^{2} \\ \displaystyle \qquad + \frac{(\lambda-\nu)(\mu-\lambda)}
      {(b-\lambda)(\lambda-c)}\, \rmd \lambda^{2} \\
    K^{ij}_{1} = \mathrm{diag}\big( 2(\nu+\lambda) g_{11}, 2(\lambda+\mu)
      g_{22}, 2(\mu+\nu) g_{33} \big) \\
    K^{ij}_{2} = \mathrm{diag}( -4\nu\lambda g_{11}, -4\lambda\mu g_{22},
      -4\mu\nu g_{33} )
  \end{array} \right. \notag \label{eq:hj:sc10}
\end{equation}\stepcounter{equation}
\begin{equation}
  \begin{array}{c} \text{Ellipsoidal:} \\ (\eta, \theta, \lambda) \\
    \text{(\theequation)} \end{array} \quad
  \left\{ \begin{array}{l}
    \displaystyle x^{2} = \frac{(a-\eta)(a-\theta)(a-\lambda)}{(a-b)(a-c)}, \:
      y^{2} = \frac{(b-\eta)(b-\theta)(b-\lambda)}{(b-a)(b-c)}, \\ \displaystyle
      \qquad z^{2} = \frac{(c-\eta)(c-\theta)(c-\lambda)}{(c-a)(c-b)} \\
    a > \eta > b > \theta > c > \lambda \\
    \displaystyle \rmd s^{2} = \frac{(\eta-\theta)(\eta-\lambda)}{4(a-\eta)
      (b-\eta)(c-\eta)}\, \rmd \eta^{2} + \frac{(\theta-\eta)(\theta-\lambda)}
      {4(a-\theta)(b-\theta)(c-\theta)}\, \rmd \theta^{2} \\ \displaystyle
      \qquad + \frac{(\lambda-\eta)(\lambda-\theta)}{4(a-\lambda)
      (b-\lambda)(c-\lambda)}\, \rmd \lambda^{2} \\
    K^{ij}_{1} = \mathrm{diag}\big( -(\theta+\lambda) g_{11},
      -(\lambda+\eta) g_{22}, -(\eta+\theta) g_{33} \big) \\
    K^{ij}_{2} = \mathrm{diag}( \theta\lambda g_{11}, \lambda\eta g_{22},
      \eta\theta g_{33} )
  \end{array} \right. \notag \label{eq:hj:sc11}
\end{equation}\stepcounter{equation}
\renewcommand{\arraystretch}{1.0}

As we are dealing with Hamiltonian systems defined in terms of
\textit{Cartesian} coordinates, the next step is to transform the components
of each of the canonical Killing tensors $K^{ij}_{1}$ and $K^{ij}_{2}$ to
Cartesian coordinates. This again is a routine calculation using the
transformations from Cartesian to separable coordinates listed in
\eref{eq:hj:sc1}--\eref{eq:hj:sc11} and the appropriate tensor transformation
law. For each of the eleven separable cases, we take a linear combination of
$K^{ij}_{1}$, $K^{ij}_{2}$ and the metric $g^{ij}$.
Using~\eref{eq:ITK2E3:Ktens_compts2}, we then identify the constants in the
linear combination and any essential parameters\footnote{These refer to any
parameters appearing in the canonical Killing tensors $K^{ij}_{1}$ and
$K^{ij}_{2}$.} with the Killing tensor parameters~\eref{eq:ITK2E3:KTparams}
appearing in \eref{eq:ITK2E3:Ktens_compts2}. We note that if the separable case
under consideration has $n$ essential parameters, then one can generally choose
$n+3$ of the Killing tensor parameters in the identification. However, in the
paraboloidal and ellipsoidal cases, it is convenient to choose more than $n+3$
parameters in the identification so that the components of the resulting
Killing tensor are polynomials in the parameters and the Cartesian coordinates.
Consequently, this leads to algebraic constraints in the Killing tensor
parameters. These constraints not only ensure that the resulting Killing tensor
has normal eigenvectors, but also guarantees that one can always (uniquely)
recover the original constants in the linear combination and all essential
parameters from the identified Killing tensor parameters. Each of the CKTs
constructed in this manner uniquely defines one of the eleven possible
\textit{orthogonal coordinate webs} in $\Eth$. We now present the results of
this procedure. For each of the eleven separable coordinate systems, we give
the components of the corresponding CKT with respect to Cartesian coordinates
and any restrictions on the Killing tensor parameters.

\begin{enumerate}
  \item Cartesian web
  \begin{equation}
    K^{ij} = \begin{pmatrix} a_{1} & 0 & 0 \\ 0 & a_{2} & 0 \\ 0 & 0 & a_{3}
      \end{pmatrix} \label{eq:hj:w1}
  \end{equation}
  \item Circular cylindrical web
  \begin{equation}
    K^{ij} = \begin{pmatrix} a_{1} + c_{3} y^{2} & -c_{3} xy & 0 \\
      -c_{3} xy & a_{1} + c_{3} x^{2} & 0 \\ 0 & 0 & a_{3} \end{pmatrix}
      \label{eq:hj:w2}
  \end{equation}
  \item Parabolic cylindrical web
  \begin{equation}
    K^{ij} = \begin{pmatrix} a_{1} \:& b_{23} y \:& 0 \\ b_{23} y \:& a_{1}
      - 2 b_{23} x \:& 0 \\ 0 \:& 0 \:& a_{3} \end{pmatrix} \label{eq:hj:w3}
  \end{equation}
  \item Elliptic-hyperbolic web
  \begin{equation}
    K^{ij} = \begin{pmatrix} a_{1} + c_{3} y^{2} & -c_{3} xy & 0 \\
      -c_{3} xy & a_{2} + c_{3} x^{2} & 0 \\ 0 & 0 & a_{3} \end{pmatrix},
      \quad \frac{a_{1}-a_{2}}{c_{3}} > 0 \label{eq:hj:w4}
  \end{equation}
  \item Spherical web
  \begin{equation}
    K^{ij} = \begin{pmatrix} a_{1} + c_{2} z^{2} + c_{3} y^{2} & -c_{3} xy
      & -c_{2} xz \\ -c_{3} xy & a_{1} + c_{3} x^{2} + c_{2} z^{2} & -c_{2}
      yz \\ -c_{2} xz & -c_{2} yz & a_{1} + c_{2} x^{2} + c_{2} y^{2}
      \end{pmatrix} \label{eq:hj:w5}
  \end{equation}
  \item Prolate spheroidal web
  \begin{equation}
    \!\! K^{ij} = \begin{pmatrix} a_{1} + c_{2} z^{2} + c_{3} y^{2} & -c_{3} xy
      & -c_{2} xz \\ -c_{3} xy & a_{1} + c_{3} x^{2} + c_{2} z^{2} & -c_{2} yz
      \\ -c_{2} xz & -c_{2} yz & a_{3} + c_{2} x^{2} + c_{2} y^{2}
      \end{pmatrix}, \quad \frac{a_{3}-a_{1}}{c_{2}} > 0 \label{eq:hj:w6}
  \end{equation}
  \item Oblate spheroidal web
  \begin{equation}
    \!\! K^{ij} = \begin{pmatrix} a_{1} + c_{2} z^{2} + c_{3} y^{2} & -c_{3} xy
      & -c_{2} xz \\ -c_{3} xy & a_{1} + c_{3} x^{2} + c_{2} z^{2} & -c_{2} yz
      \\ -c_{2} xz & -c_{2} yz & a_{3} + c_{2} x^{2} + c_{2} y^{2}
      \end{pmatrix}, \quad \frac{a_{3}-a_{1}}{c_{2}} < 0 \label{eq:hj:w7}
  \end{equation}
  \item Parabolic web
  \begin{equation}\renewcommand{\arraystretch}{1.3}
    K^{ij} = \begin{pmatrix} a_{1} - 2 b_{12} z + c_{3} y^{2} & -c_{3} xy
      & b_{12} x \\ -c_{3} xy & a_{1} - 2 b_{12} z + c_{3} x^{2} & b_{12} y \\
      b_{12} x & b_{12} y & a_{1} \end{pmatrix} \label{eq:hj:w8}
  \renewcommand{\arraystretch}{1.0}\end{equation}
  \item Conical web
  \begin{equation}
    K^{ij} = \begin{pmatrix} a_{1} + c_{2} z^{2} + c_{3} y^{2} & -c_{3} xy
      & -c_{2} zx \\ -c_{3} xy & a_{1} + c_{3} x^{2} + c_{1} z^{2} & -c_{1} yz
      \\ -c_{2} zx & -c_{1} yz & a_{1} + c_{1} y^{2} + c_{2} x^{2}
      \end{pmatrix} \label{eq:hj:w9}
  \end{equation}
  \item Paraboloidal web
  \begin{gather}\renewcommand{\arraystretch}{1.3}
    K^{ij} = \begin{pmatrix} a_{1} - 2 b_{12} z + c_{3} y^{2} & -c_{3} xy
      & b_{12} x \\ -c_{3} xy & a_{2} + 2 b_{21} z + c_{3} x^{2}
      & -b_{21} y \\ b_{12} x & -b_{21} y & a_{3} \end{pmatrix},
      \label{eq:hj:w10} \\
    b_{12}[b_{12} b_{21} + c_{3}(a_{2} - a_{3})] + b_{21}[b_{12} b_{21}
      + c_{3}(a_{1} - a_{3})] = 0 \notag
  \renewcommand{\arraystretch}{1.0}\end{gather}
  \item Ellipsoidal web
  \begin{gather}
    K^{ij} = \begin{pmatrix} a_{1} + c_{2} z^{2} + c_{3} y^{2} & -c_{3} xy
      & -c_{2} zx \\ -c_{3} xy & a_{2} + c_{3} x^{2} + c_{1} z^{2} & -c_{1} yz
      \\ -c_{2} zx & -c_{1} yz & a_{3} + c_{1} y^{2} + c_{2} x^{2}
      \end{pmatrix}, \label{eq:hj:w11} \\
    (a_{1} - a_{2}) c_{1} c_{2} + (a_{2} - a_{3}) c_{2} c_{3}
      + (a_{3} - a_{1}) c_{3} c_{1} = 0 \notag
  \end{gather}
\end{enumerate}

We remark that the eleven Killing tensors \eref{eq:hj:w1}--\eref{eq:hj:w11}
represent all possible Killing tensors with normal eigenvectors, up to
equivalence. We have essentially computed the general solution of the
TSN conditions~\eref{eq:hj:tsn} using Eisenhart's method.

As we shall see in section~\ref{sec:class}, the fundamental invariants of
$\mathcal{K}^{2}(\Eth)$ fail to discriminate amongst the eleven coordinate
webs. In anticipation, we make the following key observation. The eleven
orthogonal coordinate webs can be divided into three groups according to
table~\ref{table:webtypes}. More precisely, we say that a CKT $\vec{K} \in
\mathcal{K}^{2}(\Eth)$ is \textit{translational (rotational)} if it admits a
translational (rotational) Killing vector\footnote{The circular cylindrical
tensor~\eref{eq:hj:w2} also admits a rotational Killing vector and can
therefore be considered as both translational and rotational.} $\vec{V} \in
\mathcal{K}^{1}(\Eth)$ satisfying $\mathcal{L}_{\vec{V}} \vec{K} = 0$. This
definition still lacks complete precision for we have not defined translational
and rotational Killing vectors. Certainly, one can give a definition of such
Killing vectors in terms of integral curves. But ideally we would like to give
a definition in terms of algebraic invariants. We will revisit this problem in
section~\ref{sec:class:K1R3}. For now, it suffices to note that the canonical
translational CKTs~\eref{eq:hj:w1}--\eref{eq:hj:w4} admit the Killing vector
$\vec{V} = \vec{X}_{3}$, while the canonical rotational
CKTs~\eref{eq:hj:w5}--\eref{eq:hj:w8} admit the Killing vector $\vec{V} =
\vec{R}_{3}$.

\begin{table}[t]\begin{center}\renewcommand{\arraystretch}{1.2}
  \caption{\label{table:webtypes} The orthogonal coordinate webs in
    Euclidean space.}
  \begin{tabular}{lll} \hline\hline
    \rule[-1.5mm]{0mm}{5mm}Translational webs & Rotational webs
      & Asymmetric webs \\ \hline
    Cartesian & spherical & conical \\
    circular cylindrical & prolate spheroidal & paraboloidal \\
    parabolic cylindrical & oblate spheroidal & ellipsoidal \\
    elliptic-hyperbolic & parabolic & \\ \hline\hline
  \end{tabular}
\renewcommand{\arraystretch}{1.0}\end{center}\end{table}

We now pose the following problem: construct a subspace of $\mathcal{K}^{2}
(\Eth)$ consisting of translational Killing tensors, say with $\vec{V} =
\vec{X}_{3}$. To proceed we take the general Killing tensor from
\eref{eq:ITK2E3:Ktens} and impose the condition $\mathcal{L}_{\vec{X}_{3}}
\vec{K} = 0$. This gives a linear system of equations in the Killing tensor
parameters which can be readily solved to yield
\begin{equation}
  K^{ij} = \begin{pmatrix} a_{1} + 2 b_{13} y + c_{3} y^{2} & \alpha_{3}
    - b_{13} x + b_{23} y - c_{3} xy & \alpha_{2} - \beta_{1} y \\ \alpha_{3}
    - b_{13} x + b_{23} y - c_{3} xy & a_{2} - 2 b_{23} x + c_{3} x^{2} &
    \alpha_{1} + \beta_{1} x \\ \alpha_{2} - \beta_{1} y & \alpha_{1}
    + \beta_{1} x & a_{3} \end{pmatrix}. \label{eq:hj:Ktrans_notint}
\end{equation}
We see that the four canonical translational CKTs are all special cases of
\eref{eq:hj:Ktrans_notint}. However, it follows that the Killing tensor
\eref{eq:hj:Ktrans_notint} does not generally have normal eigenvectors.
Consequently, we cannot take our subspace to be those Killing tensors of the
form \eref{eq:hj:Ktrans_notint}. Nevertheless, using the TSN
conditions~\eref{eq:hj:tsn}, it can be shown that $\alpha_{1} = \alpha_{2} =
\beta_{1} = 0$ is a sufficient condition for \eref{eq:hj:Ktrans_notint} to have
normal eigenvectors. This is not a necessary condition, for the constant
Killing tensor $K^{ij} = A^{ij}$ is of the form \eref{eq:hj:Ktrans_notint} and
has normal eigenvectors. But, it is immediate from the transformation rules
\eref{eq:ITK2E3:trans_ABC} that a CKT is Cartesian if and only if it is a
constant Killing tensor. Moreover, it can be shown from \eref{eq:hj:tsn} that
if a Killing tensor of the form \eref{eq:hj:Ktrans_notint} has normal
eigenvectors and is \textit{not} Cartesian, then $\alpha_{1} = \alpha_{2} =
\beta_{1} = 0$. Therefore, any translational Killing tensor with orthogonally
integrable eigenvectors which is not Cartesian has the form
\begin{equation}
  K^{ij}_{T} = \begin{pmatrix} a_{1} + 2 b_{13} y + c_{3} y^{2} & \alpha_{3}
    - b_{13} x + b_{23} y - c_{3} xy & 0 \\ \alpha_{3} - b_{13} x + b_{23} y
    - c_{3} xy & a_{2} - 2 b_{23} x + c_{3} x^{2} & 0 \\ 0 & 0 & a_{3}
    \end{pmatrix}. \label{eq:hj:Ktrans}
\end{equation}
We define the subspace $\mathcal{K}_{T}^{2}(\Eth)$ of $\mathcal{K}^{2}(\Eth)$
to be the set of all Killing tensors of the form \eref{eq:hj:Ktrans}. We shall
refer to this subspace as the \textit{space of translational Killing tensors},
bearing in mind that it does not include all of the Cartesian CKTs. This does
not pose any obstacle whatsoever in our goal of constructing a classification
scheme, as we pointed out that Cartesian CKTs are trivially the constant
Killing tensors. Finally, we remark that the space $\mathcal{K}_{T}^{2}(\Eth)$
enjoys two nice features. Firstly, the form of the general translational
Killing tensor \eref{eq:hj:Ktrans} is invariant under the
(orientation-preserving) isometry group $I(\Etw)$, and secondly, the upper
$2 \times 2$ block of \eref{eq:hj:Ktrans} is the general Killing tensor on the
Euclidean plane $\mathcal{K}^{2}(\Etw)$ (see, for example, \cite{MST02}).
Consequently, we can take advantage of known results in the literature for
classifying the translational webs.

We can perform a similar analysis for the rotational webs. It follows that
any Killing tensor with normal eigenvectors admitting a Killing vector $\vec{V}
= \vec{R}_{3}$ has the form
\begin{equation}
  K^{ij}_{R} = \begin{pmatrix} a_{1} - 2 b_{12} z + c_{2} z^{2} + c_{3} y^{2}
    & -c_{3} xy & b_{12} x - c_{2} xz \\ -c_{3} xy & a_{1} - 2 b_{12} z + c_{3}
    x^{2} + c_{2} z^{2} & b_{12} y - c_{2} yz \\ b_{12} x - c_{2} xz & b_{12} y
    - c_{2} yz & a_{3} + c_{2} x^{2} + c_{2} y^{2} \end{pmatrix}.
    \label{eq:hj:Krot}
\end{equation}
We define the subspace $\mathcal{K}_{R}^{2}(\Eth)$ of $\mathcal{K}^{2}(\Eth)$
to be the set of all Killing tensors of the form \eref{eq:hj:Krot} and shall
refer to this subspace as the \textit{space of rotational Killing tensors}. We
remark that the form of the general rotational Killing tensor \eref{eq:hj:Krot}
is also invariant under the isometry group $I(\mathbb{R})$ (i.e.\ the group of
translations about the $z$-axis) and that all canonical rotational
CKTs~\eref{eq:hj:w5}--\eref{eq:hj:w8} are special cases of
\eref{eq:hj:Krot}.

In conclusion, we have defined the following four vector spaces:
$\mathcal{K}^{1}(\Eth)$, $\mathcal{K}_{T}^{2}(\Eth)$, $\mathcal{K}_{R}^{2}
(\Eth)$ and $\mathcal{K}^{2}(\Eth)$. For each of these four spaces, we need
to compute the fundamental invariants under the action of the appropriate
isometry group and classify the corresponding canonical forms. This is the
topic of the next two sections.


\section{Fundamental invariants of Killing tensors in Euclidean space}
\label{sec:inv}

In this section we derive the fundamental invariants in each of the four
vector spaces $\mathcal{K}^{1}(\Eth)$, $\mathcal{K}_{T}^{2}(\Eth)$,
$\mathcal{K}_{R}^{2}(\Eth)$ and $\mathcal{K}^{2}(\Eth)$, under the action of
their corresponding isometry group.

\subsection{The space of Killing vectors} \label{sec:inv:K1R3}

The most general Killing vector $\vec{V} \in \mathcal{K}^{1}(\Eth)$ may be
expressed as
\begin{equation}
  \vec{V} = A^{i} \vec{X}_{i} + C^{i} \vec{R}_{i}, \label{eq:inv:Kvec}
\end{equation}
where $\vec{X}_{i}$ and $\vec{R}_{i}$, $i=1,2,3$, are the Killing vectors
defined in \eref{eq:ITK2E3:Kvec} and the coefficients $A^{i}$ and $C^{i}$
are constant. For sake of convenience, we set
$$
  A^{i} = (a_{1}, a_{2}, a_{3}), \quad C^{i} = (c_{1}, c_{2}, c_{3}),
$$
so that the space of \textit{Killing vector parameters} is spanned by the six
parameters
\begin{equation}
  a_{1}, a_{2}, a_{3}, c_{1}, c_{2}, c_{3}. \label{eq:inv:KVparams}
\end{equation}
For future reference, we note that \eref{eq:inv:Kvec} can be written in the
form
\begin{equation}
  \vec{V} = (a_{1} - c_{2} z + c_{3} y) \vec{X}_{1} + (a_{2} - c_{3} x + c_{1}
    z) \vec{X}_{2} + (a_{3} - c_{1} y + c_{2} x) \vec{X}_{3} .
    \label{eq:inv:Kvec_alt}
\end{equation}
As in section~\ref{sec:ITK2E3}, we can derive the transformation rules for
the Killing vector parameters under the action of $I(\Eth)$. It follows that
\begin{equation}
  \tilde{A}^{i} = A^{j} \lambda^{i}{}_{j} + C^{j} \mu^{i}{}_{j}, \quad
  \tilde{C}^{i} = C^{j} \lambda^{i}{}_{j}, \label{eq:inv:trans_KVparams}
\end{equation}
which lead to the following infinitesimal generators of $I(\Eth)$ in the
representation defined by \eref{eq:inv:trans_KVparams}:
\begin{equation}
  \vec{U}_{i} = \epsilon^{j}{}_{ik} C^{k} \frac{\partial}{\partial A^{j}}, \quad
  \vec{V}_{i} = \epsilon^{j}{}_{ki} A^{k} \frac{\partial}{\partial A^{j}}
    + \epsilon^{j}{}_{ki} C^{k} \frac{\partial}{\partial C^{j}},
    \label{eq:inv:Kvec_UV}
\end{equation}
for $i=1,2,3$. Using the formalism of section~\ref{sec:ITKT}, it follows from
\eref{eq:inv:Kvec_UV} that $\mathcal{K}^{1}(\Eth)$ admits two fundamental
$I(\Eth)$-invariants, namely
\begin{equation}
  \Delta_{1} = C^{i} C_{i}, \quad \Delta_{2} = A^{i} C_{i}. \label{eq:inv:KVinv}
\end{equation}
In the next section, we will use these invariants to classify the elements
of $\mathcal{K}^{1}(\Eth)$ and define translational and rotational
Killing vectors in terms of $\Delta_{1}$ and $\Delta_{2}$.

\subsection{The space of translational Killing tensors} \label{sec:inv:KT2E3}

In section~\ref{sec:hj} we defined the space of translational Killing tensors
$\mathcal{K}_{T}^{2}(\Eth)$ to the set of Killing tensors of the form
\eref{eq:hj:Ktrans}. We pointed out that the upper $2 \times 2$ block of
\eref{eq:hj:Ktrans} is the general Killing tensor in the space $\mathcal{K}^{2}
(\Etw)$. The fundamental $I(\Etw)$-invariants of this vector space are known
(see, for example, \cite{MST02}), and hence are also $I(\Etw)$-invariants of
$\mathcal{K}_{T}^{2}(\Eth)$. It follows that $\mathcal{K}_{T}^{2}(\Eth)$ admits
two fundamental $I(\Etw)$-invariants given by\footnote{The paper \cite{MST02}
only treats the space of ``non-trivial'' Killing tensors in $\mathcal{K}^{2}
(\Etw)$. The invariants~\eref{eq:inv:KT2inv} form a complete set of fundamental
$I(\Etw)$-invariants for this vector subspace. It can be shown that
$\mathcal{K}_{T}^{2}(\Eth)$ admits two additional fundamental
$I(\Etw)$-invariants; we do not present them here as they play no role in
classifying the elements of $\mathcal{K}_{T}^{2}(\Eth)$.}
\begin{equation}
  \Delta_{1} = c_{3}, \quad \Delta_{2} = [b_{13}{}^{2} - b_{23}{}^{2} + c_{3}
    (a_{2} - a_{1})]^{2} + 4(b_{13} b_{23} - \alpha_{3} c_{3})^{2}
    \label{eq:inv:KT2inv}
\end{equation}
(see proposition~4.1 in \cite{MST02}).

\subsection{The space of rotational Killing tensors} \label{sec:inv:KR2E3}

The set of all Killing tensors of the form \eref{eq:hj:Krot} defines the space
of rotational Killing tensors $\mathcal{K}_{R}^{2}(\Eth)$. This subspace of
$\mathcal{K}^{2}(\Eth)$ is mapped to itself under the action of the isometry
group $I(\mathbb{R})$, the group of translations about the $z$-axis. Trivially,
the Lie algebra $i(\mathbb{R})$ is generated by the Killing vector field
$\vec{X}_{3}$ and hence the corresponding infinitesimal generator in the
parameter space is the generator $\vec{U}_{3}$ in \eref{eq:ITK2E3:gen_explicit}
restricted to $\mathcal{K}_{R}^{2}(\Eth)$ which reads
\begin{equation}
  \vec{U}_{3} = -2 b_{12} \frac{\partial}{\partial a_{1}}
    - c_{2} \frac{\partial}{\partial b_{12}} . \label{eq:inv:Krot_infgen}
\end{equation}
Solving the PDE $\vec{U}_{3}(F) = 0$ by the method of characteristics, we
obtain the four fundamental $I(\mathbb{R})$-invariants of
$\mathcal{K}_{R}^{2}(\Eth)$, namely
\begin{equation}
  \Delta_{1} = c_{2}, \quad \Delta_{2} = b_{12}{}^{2} + c_{2}(a_{3} - a_{1}),
    \quad \Delta_{3} = a_{3}, \quad \Delta_{4} = c_{3} . \label{eq:inv:KR2inv}
\end{equation}

\subsection{The space of Killing tensors} \label{sec:inv:K2E3}

We now briefly describe how to derive a complete set of invariants for the full
vector space $\mathcal{K}^{2}(\Eth)$ of valence-two Killing tensors in
Euclidean space. As mentioned at the end of section~\ref{sec:ITK2E3},
$\mathcal{K}^{2}(\Eth)$ admits fifteen fundamental $I(\Eth)$-invariants which
can be computed by solving the system of linear PDEs
\begin{equation}
  \vec{U}_{i}(F) = 0, \quad \vec{V}_{i}(F) = 0, \quad i=1,2,3,
    \label{eq:inv:PDE}
\end{equation}
where the generators $\vec{U}_{i}$ and $\vec{V}_{i}$ are given by
\eref{eq:ITK2E3:gen_explicit} and $F$ is an analytic function in the Killing
tensor parameters.

Computationally, arriving at the general solution of \eref{eq:inv:PDE} is
non-trivial. In particular, we have found that the method of characteristics
becomes intractable when applied to \eref{eq:inv:PDE}. However, we have
successfully computed all fifteen invariants using the method of undetermined
coefficients \cite{DHMS04}. The simplest implementation of this method is to
build monomial trial functions in the Killing tensor parameters up to a fixed
degree, take a linear combination of these monomials and substitute the
combination into \eref{eq:inv:PDE} leading to a large (sparse) system of linear
equations in the undetermined coefficients. This approach has two obvious
disadvantages. Firstly, such an ansatz does not take advantage of the apparent
structure and symmetry of \eref{eq:inv:PDE}. Consequently, as one must
construct all monomials up to and including degree five to recover all fifteen
fundamental invariants, approximately $50\,000$ undetermined coefficients are
involved; the corresponding linear system requires almost ninety hours of CPU
time to solve on a modest Sun workstation! Secondly, by virtue of the ansatz,
the computed invariants are in expanded form and occupy fifteen pages of
output.  A more effective ansatz involves constructing scalar trial functions
which are ``tensorial''\footnote{As the $A^{ij}$ and $B^{ij}$ do not transform
as tensors (see \eref{eq:ITK2E3:trans_ABC}), there is no reason why this ansatz
should work.} in the $A^{ij}$, $B^{ij}$ and $C^{ij}$. For example, trial
functions which are cubic in the $C^{ij}$ include
$$
  C_{i}{}^{i} C_{j}{}^{j} C_{k}{}^{k}, \quad C^{ij} C_{ij} C_{k}{}^{k}, \quad
  C^{ij} C_{j}{}^{k} C_{ki}, \quad \epsilon_{ikm} \epsilon_{j\ell n} C^{ij}
  C^{k\ell} C^{mn} .
$$
Implementing the method of undetermined coefficients with this ansatz yields
the following fifteen fundamental $I(\Eth)$-invariants of $\mathcal{K}^{2}
(\Eth)$:
\begin{align}\begin{split}
  \Delta_{1} &= B_{i}{}^{i}, \:\: \Delta_{2} = C_{i}{}^{i}, \:\:
    \Delta_{3} = B^{ij} C_{ij}, \:\: \Delta_{4} = C^{ij} C_{ij}, \:\:
    \Delta_{5} = B^{ij} B_{ji} + A^{ij} C_{ij}, \\
  \Delta_{6} &= B^{ij} C_{j}{}^{k} C_{ki}, \quad \Delta_{7} = C^{ij}
    C_{j}{}^{k} C_{ki}, \quad \Delta_{8} = C^{ij} [ B_{j}{}^{k} ( B_{ik}
    + 2 B_{ki} ) + A_{j}{}^{k} C_{ki} ],\\
  \Delta_{9} &= \epsilon_{ikm} \epsilon_{j\ell n} B^{ij} B^{k\ell} B^{mn}
    - 2 ( B_{i}{}^{[i} B_{j}{}^{j]} + A^{ij} C_{ij} ) B_{k}{}^{k}
    + 6 B^{ij} A_{j}{}^{k} C_{ki}, \\
  \Delta_{10} &= B^{ij} ( B_{i}{}^{k} C_{kj} - 2 B_{j}{}^{k} C_{ki} )
    - ( B^{ij} B_{ij} + A^{ij} C_{ij} ) C_{k}{}^{k} + A_{i}{}^{i}
    C_{j}{}^{[j} C_{k}{}^{k]}, \\
  \Delta_{11} &= \epsilon_{i\ell m} \epsilon_{jkp} B^{ij} B^{k\ell} C^{mn}
    C_{n}{}^{p} + B^{ij} [ B_{ij} C^{k\ell} C_{k\ell} - C_{j}{}^{k} (
    C_{k}{}^{\ell} B_{i\ell} + 4 C_{[k}{}^{\ell} B_{\ell]i} ) ] \\
  &\qquad + A^{ij} C_{ij} C_{k}{}^{[k} C_{\ell}{}^{\ell]}, \\
  \Delta_{12} &= A_{i}{}^{i} [ ( C_{j}{}^{j} C_{k}{}^{k} + 3 C^{jk} C_{jk} )
    C_{\ell}{}^{\ell} - 4 C^{jk} C_{k}{}^{\ell} C_{\ell j} ] - 6 A^{ij} C_{ij}
    C^{k\ell} C_{k\ell} \\
  &\qquad + 6 B^{ij} \{ B_{ij} C^{k\ell} C_{k\ell} - C_{j}{}^{k} [
    ( B_{ik} - 2 B_{ki}) C_{\ell}{}^{\ell} + 4 C_{k}{}^{\ell} B_{\ell i}]\} \\
  &\qquad + 12 \epsilon_{i\ell m} \epsilon_{jkp} B^{ij} B^{k\ell} C^{mn}
    C_{n}{}^{p}, \\
  \Delta_{13} &= A^{ij} ( B_{ij} C_{k}{}^{[k} C_{\ell}{}^{\ell]} + B_{j}{}^{k}
    C_{k}{}^{\ell} C_{\ell i} - 2 C_{i(j} B_{k)}{}^{k} C_{\ell}{}^{\ell} ) \\
  &\qquad + A_{i}{}^{i} C^{jk} ( B_{jk} C_{\ell}{}^{\ell} - B_{k}{}^{\ell}
    C_{\ell j} ) - B^{ij} [ B_{ij} B^{k\ell} C_{k\ell} + 2 C_{j}{}^{k} B_{ki}
    B_{\ell}{}^{\ell} \\
  &\qquad + B_{j}{}^{k} B_{ik} C_{\ell}{}^{\ell} - ( B_{j}{}^{k} B_{i\ell}
    + B_{i}{}^{k} B_{\ell j} ) C_{k}{}^{\ell} ], \\
  \Delta_{14} &= 4 A_{i}{}^{[i} A_{j}{}^{j]} C_{k}{}^{[k} C_{\ell}{}^{\ell]}
    + 8 A^{ij} ( A_{j}{}^{k} C_{k[i} C_{\ell]}{}^{\ell} + A_{k}{}^{k}
    C_{[j}{}^{\ell} C_{\ell]i} ) + A^{ij} C_{ij} ( A^{k\ell} C_{k\ell} \\
  &\qquad + 4 B_{k}{}^{[k} B_{\ell}{}^{\ell]} ) + 4 C^{ij} B_{j}{}^{k}
    A_{k}{}^{\ell} B_{i\ell} + 16 A^{ij} C_{j}{}^{k} B_{[k}{}^{\ell}
    B_{\ell]i}, \\
  \Delta_{15} &= A^{ij} C_{ij} [ ( C_{k}{}^{k} C_{\ell}{}^{\ell} - 3 C^{k\ell}
    C_{k\ell} ) C_{m}{}^{m} + 2 C^{k\ell} C_{\ell}{}^{m} C_{mk} ] \\
  &\qquad - 6 A^{ij} C_{j}{}^{k} C_{ki} C_{\ell}{}^{[\ell} C_{m}{}^{m]}
    - 12 C^{ij} B_{j}{}^{k} ( C_{k}{}^{\ell} B_{i[\ell} C_{m]}{}^{m}
    + 2 B_{k}{}^{\ell} C_{i[\ell} C_{m]}{}^{m} ).
\end{split}\label{eq:inv:K2inv}\end{align}
We have hence proven the following theorem.

\begin{theorem}
  Consider the vector space $\mathcal{K}^{2}(\Eth)$. Any algebraic
  $I(\Eth)$-invariant $\mathcal{I}$ defined over $\mathcal{K}^{2}(\Eth)$ in
  terms of the Killing tensor parameters~\eref{eq:ITK2E3:KTparams} where the
  isometry group $I(\Eth)$ acts freely and regularly with six-dimensional
  orbits can be locally uniquely expressed as an analytic function
  $\mathcal{I} = F(\Delta_{1}, \ldots, \Delta_{15})$, where the fundamental
  $I(\Eth)$-invariants $\Delta_{i}$, $i=1,\ldots,15$, are given by
  \eref{eq:inv:K2inv}.\label{theorem:inv}
\end{theorem} 

We close this section by addressing a minor computational issue. It is clear
from \eref{eq:ITK2E3:Ktens_compts2} that the Killing tensor parameters
$b_{11}$, $b_{22}$ and $b_{33}$ are \textit{not} uniquely determined for a
given Killing tensor $\vec{K} \in \mathcal{K}^{2}(\Eth)$. Thus, how does one
evaluate the invariants~\eref{eq:inv:K2inv} on a given Killing tensor? Indeed,
we require an invariant method for solving the equations \eref{eq:ITK2E3:beta}
for the $b_{ii}$ in terms of the known $\beta_{i}$. This problem is easily
rectified upon observing that the equations \eref{eq:ITK2E3:beta} in
conjunction with the condition $\Delta_{1} = 0$ $\Leftrightarrow$ $b_{11} +
b_{22} + b_{33} = 0$ yields a unique solution for the $b_{ii}$ given by
\begin{equation}
  b_{11} = \tfrac{1}{3}(\beta_{3} - \beta_{2}), \quad
  b_{22} = \tfrac{1}{3}(\beta_{1} - \beta_{3}), \quad
  b_{33} = \tfrac{1}{3}(\beta_{2} - \beta_{1}). \label{eq:inv:bii}
\end{equation}
In what follows, we shall extract the parameters $b_{11}$, $b_{22}$ and
$b_{33}$ from a given $\vec{K} \in \mathcal{K}^{2}(\Eth)$ using
\eref{eq:inv:bii}.


\section{Invariant classification of orthogonal coordinate webs in Euclidean
  space} \label{sec:class}

In order to build a classification scheme for the orthogonal coordinate webs in
Euclidean space based on the set of Killing tensors in $\mathcal{K}^{2}(\Eth)$
with normal eigenvectors and distinct eigenvalues, one must first know how to
classify elements in the vector spaces $\mathcal{K}^{1}(\Eth)$,
$\mathcal{K}_{T}^{2}(\Eth)$ and $\mathcal{K}_{R}^{2}(\Eth)$. The classification
of elements in these spaces are treated in subsections~\ref{sec:class:K1R3},
\ref{sec:class:KT2E3} and \ref{sec:class:KR2E3}, respectively. We then use
these results to classify the orthogonal coordinate webs of
$\mathcal{K}^{2}(\Eth)$ in subsection~\ref{sec:class:K2E3}.

\subsection{Classification of Killing vectors} \label{sec:class:K1R3}

We shall classify the elements of $\mathcal{K}^{1}(\Eth)$ according to whether
the fundamental invariants~\eref{eq:inv:KVinv} are zero or non-zero. There are
three cases to consider each of which gives rise to a canonical Killing vector.
The classification scheme is summarized in table~\ref{table:class:K1R3}

\begin{table}[t]\begin{center}\renewcommand{\arraystretch}{1.2}
  \caption{\label{table:class:K1R3} Invariant classification of Killing vectors
    in Euclidean space.}
  \begin{tabular}{lll} \hline\hline
    \rule[-1.5mm]{0mm}{5mm}Classification & \multicolumn{1}{l}{Invariants}
      \\ \hline
    translational & $\Delta_{1} = 0$, & $\Delta_{2} = 0$ \\
    rotational & $\Delta_{1} \not= 0$, & $\Delta_{2} = 0$ \\
    helicoidal & $\Delta_{1} \not= 0$, & $\Delta_{2} \not= 0$ \\ \hline\hline
  \end{tabular}
\renewcommand{\arraystretch}{1.0}\end{center}\end{table}

\begin{enumerate}
  \item $\Delta_{1} = 0$, $\Delta_{2} = 0$. In this case, the Killing vector
    $\vec{V}$ is of the form
    \begin{equation}
      \vec{V} = A^{i} \vec{X}_{i}, \label{eq:class:K1R3a}
    \end{equation}
    where it is assumed that the $A^{i}$ are not all zero so that $\vec{V}$ is
    non-trivial. It follows that we can use the isometry group $I(\Eth)$ to
    transform \eref{eq:class:K1R3a} to
    \begin{equation}
      \vec{\tilde{V}} = \tilde{a}_{3} \vec{\tilde{X}}_{3},
        \label{eq:class:K1R3b}
    \end{equation}
    for some $\tilde{a}_{3} \not= 0$. Indeed, the transformation rules
    \eref{eq:inv:trans_KVparams} reduce to $\tilde{C}^{i} = 0$ and
    $\tilde{A}^{i} = A^{j} \lambda^{i}{}_{j}$. Without loss of generality, we
    can set the components of the translation $\delta^{i} = 0$. The components
    of the rotation $\lambda_{j}{}^{i}$ can be computed by setting
    $\lambda^{3}{}_{j} = (A^{k} A_{k})^{-1/2} A_{j}$ and then obtaining
    $\lambda^{1}{}_{j}$ and $\lambda^{2}{}_{j}$ by extending the vector
    $\lambda^{3}{}_{j}$ to a proper orthonormal basis in $\Eth$ (using the
    Gram-Schmidt algorithm or QR decomposition, for example). We observe that
    \eref{eq:class:K1R3b} defines a translation, hence we say that a Killing
    vector $\vec{V} \in \mathcal{K}^{1}(\Eth)$ is \textit{translational} iff
    $\Delta_{1} = \Delta_{2} = 0$.\medskip
    
  \item $\Delta_{1} \not= 0$, $\Delta_{2} = 0$. Using the isometry group
    $I(\Eth)$, we claim that any such Killing vector $\vec{V}$ can be
    transformed to
    \begin{equation}
      \vec{\tilde{V}} = \tilde{c}_{3} \vec{\tilde{R}}_{3},
        \label{eq:class:K1R3c}
    \end{equation}
    for some $\tilde{c}_{3} \not= 0$. Indeed, we can first make a translation
    $x^{i} = \hat{x}^{i} + \delta^{i}$ so that $\hat{A}^{i} = 0$ in the new
    coordinates $\hat{x}^{i}$. It follows from the transformation rules
    \eref{eq:inv:trans_KVparams} and \eref{eq:ITK2E3:mu} that the $\delta^{i}$
    must satisfy the system of linear equations $\epsilon^{i}{}_{jk} C^{j}
    \delta^{k} = A^{i}$, which has a solution iff $\Delta_{2} = 0$. In the new
    coordinates $\hat{x}^{i}$, it follows that $\vec{\hat{V}} = C^{i}
    \vec{\hat{R}}_{i}$. By a similar argument to that used in the previous
    case, we can find a rotation $\lambda_{j}{}^{i}$ such the coordinate
    transformation $\hat{x}^{i} = \lambda_{j}{}^{i} \tilde{x}^{j}$ puts
    $\vec{\hat{V}}$ into the form~\eref{eq:class:K1R3c}. We observe that
    \eref{eq:class:K1R3c} defines a rotation, hence we say that a Killing
    vector $\vec{V} \in \mathcal{K}^{1}(\Eth)$ is \textit{rotational} iff
    $\Delta_{1} \not= 0$ and $\Delta_{2} = 0$.\medskip
    
  \item $\Delta_{1} \not= 0$, $\Delta_{2} \not= 0$. By similar arguments to
    those used in the previous two cases, it can be shown in this case that
    there exists a coordinate transformation \eref{eq:ITK2E3:trans_x} such that
    \begin{equation}
      \vec{\tilde{V}} = \tilde{a}_{3} \vec{\tilde{X}}_{3} + \tilde{c}_{3}
        \vec{\tilde{R}}_{3} , \label{eq:class:K1R3d}
    \end{equation}
    for some non-zero $\tilde{a}_{3}$ and $\tilde{c}_{3}$. The integral curves
    of such a Killing vector field are helices and so we say that a Killing
    vector $\vec{V} \in \mathcal{K}^{1}(\Eth)$ is \textit{helicoidal} iff
    both $\Delta_{1} \not= 0$ and $\Delta_{2} \not= 0$. We remark that we
    can further refine this classification into left and right-handed
    helicoidal Killing vectors according to the sign of $\Delta_{2}$.
\end{enumerate}

\subsection{The space of translational Killing tensors} \label{sec:class:KT2E3}

A classification scheme for the vector space of translational Killing tensors
$\mathcal{K}_{T}^{2}(\Eth)$ based on the fundamental
$I(\Etw)$-invariants~\eref{eq:inv:KT2inv} is provided in \cite{MST02}. For
completeness, we summarize this scheme in table~\ref{table:class:KT2E3}.

\begin{table}[t]\begin{center}\renewcommand{\arraystretch}{1.2}
  \caption{\label{table:class:KT2E3} Invariant classification of translational
    Killing tensors in Euclidean space.}
  \begin{tabular}{lll} \hline\hline
    \rule[-1.5mm]{0mm}{5mm}Orthogonal coordinate web
      & \multicolumn{2}{l}{Invariants} \\ \hline
    Cartesian & $\Delta_{1} = 0$, & $\Delta_{2} = 0$ \\
    circular cylindrical & $\Delta_{1} \not= 0$, & $\Delta_{2} = 0$ \\
    parabolic cylindrical & $\Delta_{1} = 0$, & $\Delta_{2} \not= 0$ \\
    elliptic-hyperbolic & $\Delta_{1} \not= 0$, & $\Delta_{2} \not= 0$
      \\ \hline\hline
  \end{tabular}
\renewcommand{\arraystretch}{1.0}\end{center}\end{table}

\subsection{The space of rotational Killing tensors} \label{sec:class:KR2E3}

Evaluating the $I(\mathbb{R})$-invariants \eref{eq:inv:KR2inv} on each
of the rotational CKTs \eref{eq:hj:w5}--\eref{eq:hj:w8} produces a
classification scheme for the vector space $\mathcal{K}_{R}^{2}(\Eth)$. It
turns out that we only need to use the invariants $\Delta_{1}$ and $\Delta_{2}$
in \eref{eq:inv:KR2inv} to obtain a classification. As we pointed out in
section~\ref{sec:hj}, we can also include the circular cylindrical web in this
classification. Our classification scheme is detailed in
table~\ref{table:class:KR2E3}.

\begin{table}[t]\begin{center}\renewcommand{\arraystretch}{1.2}
  \caption{\label{table:class:KR2E3} Invariant classification of rotational
    Killing tensors in Euclidean space.}
  \begin{tabular}{lll} \hline\hline
    \rule[-1.5mm]{0mm}{5mm}Orthogonal coordinate web
      & \multicolumn{2}{l}{Invariants} \\ \hline
    circular cylindrical & $\Delta_{1} = 0$, & $\Delta_{2} = 0$ \\
    spherical & $\Delta_{1} \not= 0$, & $\Delta_{2} = 0$ \\
    prolate spheroidal & $\Delta_{1} \not= 0$, & $\Delta_{2} > 0$ \\
    oblate spheroidal & $\Delta_{1} \not= 0$, & $\Delta_{2} < 0$ \\
    parabolic & $\Delta_{1} = 0$, & $\Delta_{2} \not= 0$ \\ \hline\hline
  \end{tabular}
\renewcommand{\arraystretch}{1.0}\end{center}\end{table}

\subsection{The space of Killing tensors} \label{sec:class:K2E3}

The motivation for constructing invariant classification schemes in the
vector spaces treat\-ed in the previous three subsections is due to the fact
that we have been unable to obtain a scheme based solely on the fifteen
fundamental $I(\Eth)$-invariants of the full space $\mathcal{K}^{2}(\Eth)$
presented in \eref{eq:inv:K2inv}. This is primarily because these invariants
fail to discriminate amongst some of the canonical CKTs
\eref{eq:hj:w1}--\eref{eq:hj:w11}.

To begin, let $\vec{K} \in \mathcal{K}^{2}(\Eth)$ be the given CKT for which we
wish to classify. As we showed in section~\ref{sec:hj}, if $\vec{K}$ is
constant, then it necessarily characterizes a Cartesian web. Let us therefore
assume for the remainder of this section that $\vec{K}$ is not constant. The
classification of $\vec{K}$ involves two main steps:
\begin{enumerate}
  \item Determine whether $\vec{K}$ characterizes a translational, rotational
    or asymmetric web (according to the type of Killing vector it admits).
  \item Use the classification schemes in tables~\ref{table:class:KT2E3}
    and \ref{table:class:KR2E3} if $\vec{K}$ is translational or rotational,
    or, the classification scheme outlined in table~\ref{table:class:K2E3} if
    $\vec{K}$ characterizes an asymmetric web (i.e.\ $\vec{K}$ admits no
    Killing vector).
\end{enumerate}

To proceed with the first step, we let $\vec{V}$ be the general Killing vector
from \eref{eq:inv:Kvec_alt} and impose the condition
\begin{equation}
  \mathcal{L}_{\vec{V}} \vec{K} = 0 . \label{eq:class:K2E3a}
\end{equation}
Equation~\eref{eq:class:K2E3a} results in a linear system of equations in the
six Killing vector parameters~\eref{eq:inv:KVparams} which can be readily
solved. It follows that the general solution of \eref{eq:class:K2E3a} can be
decomposed as
$$
  \vec{V} = \ell_{1} \vec{V}_{1} + \cdots \ell_{n} \vec{V}_{n},
$$
for some $n \leq 6$, where $\ell_{i}$, $i = 1, \ldots, n$, are arbitrary
non-zero constants and $\{\vec{V}_{1}, \ldots,$ $\vec{V}_{n}\}$ is a linearly
independent set of Killing vectors. If $n = 0$, we conclude that
$\vec{K}$ does not admit a Killing vector, and hence characterizes an
asymmetric web. Otherwise, using table~\ref{table:class:K1R3}, we classify
each of the $\vec{V}_{i}$ according to whether they are translational,
rotational, or helicoidal. Therefore, if one of the $\vec{V}_{i}$ is
translational, then $\vec{K}$ characterizes a translational web, otherwise
$\vec{K}$ characterizes a rotational web\footnote{It is impossible for all of
the $\vec{V}_{i}$ to be helicoidal. Although the circular cylindrical web is
the only coordinate web which admits a helicoidal Killing vector, it also
admits both translational and rotational Killing vectors. Clearly, if one of
the $\vec{V}_{i}$ is helicoidal, we can conclude immediately that $\vec{K}$
characterizes a circular cylindrical web.}.

We have now shown how to determine if $\vec{K}$ characterizes a translational,
rotational or asymmetric web. To proceed, suppose that $\vec{K}$ is a
translational (rotational) Killing tensor and let $\vec{V}$ be its
corresponding translational (rotational) Killing vector. From the results of
subsection~\ref{sec:class:KT2E3} (\ref{sec:class:KR2E3}), we can use the
isometry group $I(\Eth)$ to bring $\vec{V}$ to the canonical form
$\tilde{a}_{3} \vec{\tilde{X}}_{3}$ ($\tilde{c}_{3} \vec{\tilde{R}}_{3}$).
Applying the corresponding coordinate transformation to the Killing tensor
$\vec{K}$ places it in the subspace $\mathcal{K}_{T}^{2}(\Eth)$
($\mathcal{K}_{R}^{2}(\Eth)$). Finally, we can classify the transformed
$\vec{K}$ using table~\ref{table:class:KT2E3} (\ref{table:class:KR2E3}).

Suppose now that $\vec{K}$ characterizes an asymmetric web. Using the
fundamental $I(\Eth)$-invariants~\eref{eq:inv:K2inv} of
$\mathcal{K}^{2}(\Eth)$, we shall derive a scheme for classifying $\vec{K}$. To
begin, we evaluate the invariants $\Delta_{2}$, $\Delta_{4}$ and $\Delta_{7}$
(see~\eref{eq:inv:K2inv}) on the three asymmetric CKTs
\eref{eq:hj:w9}--\eref{eq:hj:w11}. It follows that for the conical and
ellipsoidal tensors, $(\Delta_{2}, \Delta_{4}, \Delta_{7})$ $= (c_{1} + c_{2}
+ c_{3}, c_{1}{}^{2} + c_{2}{}^{2} + c_{3}{}^2, c_{1}{}^{3} + c_{2}{}^{3} +
c_{3}{}^3)$, while for the paraboloidal tensor, $(\Delta_{2}, \Delta_{4},
\Delta_{7}) = (c_{3}, c_{3}{}^2, c_{3}{}^{3})$. This motivates defining two
auxiliary invariants
\begin{equation}
  \Xi_{1} = \Delta_{2}{}^{2} - \Delta_{4}, \quad
  \Xi_{2} = \Delta_{2}{}^{3} - \Delta_{7}, \label{eq:class:K2E3_Xi12}
\end{equation}
noting that $\Xi_{1} = \Xi_{2} = 0$ on the paraboloidal
tensor~\eref{eq:hj:w10}. We claim that the vanishing of $\Xi_{1}$ and $\Xi_{2}$
is also a sufficient condition for $\vec{K}$ to characterize a paraboloidal
web. Indeed, it follows that $\Xi_{1} = \Xi_{2} = 0$ in one or more of the
following three cases
$$
  c_{1} = c_{2} = 0, \quad c_{2} = c_{3} = 0, \quad c_{3} = c_{1} = 0.
$$
The conical case~\eref{eq:hj:w9} and the ellipsoidal case~\eref{eq:hj:w11}
cannot have $c_{1} = c_{2} = 0$, since this condition reduces them to the
spherical and elliptic-hyperbolic tensors, respectively. Moreover, by a
rotation, the two other cases are also impossible for conical and ellipsoidal
tensors. Therefore, we conclude that $\vec{K}$ characterizes a paraboloidal web
if and only if $\Xi_{1} = \Xi_{2} = 0$.

Suppose now that $\vec{K}$ does not characterize a paraboloidal web for the
remainder of this section. It is convenient to define
\begin{equation}
  \Xi_{3} = 3 \Delta_{4} - \Delta_{2}{}^{2}, \label{eq:class:K2E3_Xi3}
\end{equation}
noting that $\Xi_{3} = (c_{1} - c_{2})^{2} + (c_{2} - c_{3})^{2} + (c_{3} -
c_{1})^{2}$ on both the conical and ellipsoidal CKTs \eref{eq:hj:w9} and
\eref{eq:hj:w11}. Indeed, if $\Xi_{3} = 0$, the web cannot possibly be a
conical tensor since $c_{1} = c_{2} = c_{3}$ reduces it to a special case of
the spherical tensor. Therefore, if $\Xi_{3} = 0$, then $\vec{K}$ necessarily
characterizes an ellipsoidal web.

To distinguish between a conical and ellipsoidal tensor in the case when
$\Xi_{3} \not= 0$, we define three additional auxiliary invariants given by
\begin{align}\begin{split}
  \Xi_{4} &= \Delta_{2} \Delta_{5} - 3 \Delta_{8} - 2 \Delta_{10}, \\
  \Xi_{5} &= \Delta_{2} \Delta_{10} + \Delta_{4} \Delta_{5} - \Delta_{11}, \\
  \Xi_{6} &= \Delta_{2} [ 2 \Delta_{2} ( 10 \Delta_{2} \Delta_{5}
    + 24 \Delta_{8} - 3 \Delta_{10} ) - 72 \Delta_{11} + \Delta_{12} ] \\
  &\qquad - 48 \Delta_{4} \Delta_{8} - 20 \Delta_{5} \Delta_{7} + 16 \Delta_{15}
\end{split}\label{eq:class:K2E3_Xi456}\end{align}
It follows that the invariants~\eref{eq:class:K2E3_Xi456} all evaluate to zero
on the conical tensor~\eref{eq:hj:w9}. We claim that $\Xi_{4} = \Xi_{5} =
\Xi_{6} = 0$ (in conjunction with $\Xi_{3} \not= 0$) is also a sufficient
condition for a conical tensor. Indeed, arguing by contradiction, it follows
that
\begin{align*}
  \Xi_{4} &= ( a_{1} + a_{2} - 2 a_{3} ) c_{1} c_{2} + ( a_{2} + a_{3}
    - 2 a_{1} ) c_{2} c_{3} + ( a_{3} + a_{1} - 2 a_{2} ) c_{3} c_{1}, \\
  \Xi_{5} &= ( c_{1} c_{2} + c_{2} c_{3} + c_{3} c_{1} ) [ ( a_{1} + a_{2}
    - 2 a_{3} ) c_{3} + ( a_{2} + a_{3} - 2 a_{1} ) c_{1} \\
    &\qquad + ( a_{3} + a_{1} - 2 a_{2} ) c_{2} ], \\
  \Xi_{6} &= 12 c_{1} c_{2} c_{3} [ ( 2 a_{1} - a_{2} - a_{3} ) c_{1}
    + ( 2 a_{2} - a_{3} - a_{1} ) c_{2} + ( 2 a_{3} - a_{1} - a_{2} ) c_{3} ],
\end{align*}
on the ellipsoidal tensor~\eref{eq:hj:w11} and that $\Xi_{4} = \Xi_{5} =
\Xi_{6} = 0$ in one or more of the following five cases\footnote{This
calculation is facilitated by use of the Maple Gr\"{o}bner basis package.}:
$$
  c_{1} = c_{2} = 0, \quad c_{2} = c_{3} = 0, \quad c_{3} = c_{1} = 0, \quad
  c_{1} = c_{2} = c_{3}, \quad a_{1} = a_{2} = a_{3} .
$$
By previous arguments, the first three cases are impossible if the tensor is
ellipsoidal. The fourth case can also be eliminated since $\Xi_{3} \not= 0$.
Finally, the fifth case is impossible since it reduces to a conical tensor.
Therefore, we conclude that if $\Xi_{3} \not= 0$, then the tensor is conical
if and only if $\Xi_{4} = \Xi_{5} = \Xi_{6} = 0$.

This completes the derivation of the classification scheme for the set of
asymmetric CKTs in Euclidean space. These results are summarized in
table~\ref{table:class:K2E3}.

\begin{table}[t]\begin{center}\renewcommand{\arraystretch}{1.2}
  \caption{\label{table:class:K2E3} Invariant classification of asymmetric
    Killing tensors in Euclidean space.}
  \begin{tabular}{llll} \hline\hline
    \rule[-1.5mm]{0mm}{5mm}Orthogonal coordinate web
      & \multicolumn{3}{l}{Invariants} \\ \hline
      paraboloidal & $(\Xi_{1}, \Xi_{2}) = (0,0)$ && \\
      ellipsoidal & $(\Xi_{1}, \Xi_{2}) \not= (0,0)$, & $\Xi_{3} = 0$ & or \\
      & $(\Xi_{1}, \Xi_{2}) \not= (0,0)$, & $\Xi_{3} \not= 0$,
        & $(\Xi_{4}, \Xi_{5}, \Xi_{6}) \not= (0,0,0)$ \\
      conical & $(\Xi_{1}, \Xi_{2}) \not= (0,0)$, & $\Xi_{3} \not= 0$,
        & $(\Xi_{4}, \Xi_{5}, \Xi_{6}) = (0,0,0)$ \\ \hline\hline
  \end{tabular}
\renewcommand{\arraystretch}{1.0}\end{center}\end{table}


\section{Transformations to canonical form} \label{sec:cform}

Once a CKT in Euclidean space has been classified using the scheme detailed in
the previous section, the isometry group $I(\Eth)$ can be used to transform the
Killing tensor into its corresponding canonical form. This step leads directly
to the transformation to separable coordinates. In this section we provide
methods for determining the transformations to canonical form. As we shall see,
the majority of the calculations amount to elementary linear algebra.

The procedure can be summarized as follows. Suppose that $K^{ij}$ are the
components of a CKT with respect to Cartesian coordinates $x^{i}$. Under the 
action of $I(\Eth)$, the transformation from the original set of Cartesian
coordinates $x^{i}$ to another set $\tilde{x}^{i}$ is given by
\eref{eq:ITK2E3:trans_x}, i.e.
\begin{equation}
  x^{i} = \lambda_{j}{}^{i}\tilde{x}^{j} + \delta^{i}. \label{eq:cform:trans_x}
\end{equation}
Thus, we need to determine the rotation $\lambda_{j}{}^{i} \in SO(3)$ and
translation $\delta^{i} \in \mathbb{R}^{3}$ which brings $K^{ij}$ to its
appropriate canonical form $\tilde{K}^{ij}$ given by one of the eleven
cases \eref{eq:hj:w1}-\eref{eq:hj:w11} (in the coordinates $\tilde{x}^{i}$).
Moreover, any essential parameters appearing in the tensor also need to be
determined (e.g. the parameter $a$ appearing in the elliptic-hyperbolic tensor
listed in \eref{eq:hj:sc4}). Once $\lambda_{j}{}^{i}$, $\delta^{i}$ and all
essential parameters are known, the transformation from Cartesian coordinates
$x^{i}$ to separable coordinates $u^{i}$ is
\begin{equation}
  x^{i} = \lambda_{j}{}^{i} T^{j}(u^{k}) + \delta^{i},
    \label{eq:cform:trans_sep}
\end{equation}
where $x^{i} = T^{i}(u^{j})$ is the standard coordinate transformation
associated with the separable coordinates tabulated in
\eref{eq:hj:sc1}--\eref{eq:hj:sc11}.

To carry out this procedure, we use the transformation rules relating the
parameter matrices of $K^{ij}$ to those of its canonical form $\tilde{K}^{ij}$
(see~\eref{eq:ITK2E3:ABCmatrices}). In matrix form, these transformation
rules read
\begin{subequations}\begin{align}
  \vec{\tilde{A}} &= \vec{\lambda}^{t} \vec{A} \vec{\lambda} + 2\, \mathcal{S}
    (\vec{\lambda}^{t} \vec{B} \vec{\mu}) + \vec{\mu}^{t} \vec{C} \vec{\mu},
    \label{eq:cform:paramtransA} \\
  \vec{\tilde{B}} &= \vec{\lambda}^{t} \vec{B} \vec{\lambda} + \vec{\mu}^{t}
    \vec{C} \vec{\lambda}, \label{eq:cform:paramtransB} \\
  \vec{\tilde{C}} &= \vec{\lambda}^{t} \vec{C} \vec{\lambda},
    \label{eq:cform:paramtransC}
\end{align}\label{eq:cform:paramtrans}\end{subequations}
where $\vec{x} = \vec{\lambda} \vec{\tilde{x}} + \vec{\delta}$, $\mathcal{S}$
denotes the symmetric part and
$$
  \vec{x} = \begin{pmatrix} x \\ y \\ z \end{pmatrix}_{i} = x^{i}, \quad
  \vec{\tilde{x}} = \begin{pmatrix} \tilde{x} \\ \tilde{y} \\ \tilde{z}
    \end{pmatrix}_{i} = \tilde{x}^{i}, \quad
  \vec{\lambda} = \begin{pmatrix} \lambda_{1}{}^{1} & \lambda_{2}{}^{1}
    & \lambda_{3}{}^{1} \\ \lambda_{1}{}^{2} & \lambda_{2}{}^{2}
    & \lambda_{3}{}^{2} \\ \lambda_{1}{}^{3} & \lambda_{2}{}^{3}
    & \lambda_{3}{}^{3} \end{pmatrix}_{ij} = \lambda_{j}{}^{i},
$$
$$
  \vec{\delta} = \begin{pmatrix} \delta^{1} \\ \delta^{2} \\ \delta^{3}
    \end{pmatrix}_{i} = \delta^{i}, \:\:
  \vec{\mu} = \begin{pmatrix}
    \lambda_{1}{}^{2} \delta^{3} - \lambda_{1}{}^{3} \delta^{2} \:&
    \lambda_{2}{}^{2} \delta^{3} - \lambda_{2}{}^{3} \delta^{2} \:&
    \lambda_{3}{}^{2} \delta^{3} - \lambda_{3}{}^{3} \delta^{2} \\
    \lambda_{1}{}^{3} \delta^{1} - \lambda_{1}{}^{1} \delta^{3} \:&
    \lambda_{2}{}^{3} \delta^{1} - \lambda_{2}{}^{1} \delta^{3} \:&
    \lambda_{3}{}^{3} \delta^{1} - \lambda_{3}{}^{3} \delta^{1} \\
    \lambda_{1}{}^{1} \delta^{2} - \lambda_{1}{}^{2} \delta^{1} \:&
    \lambda_{2}{}^{1} \delta^{2} - \lambda_{2}{}^{2} \delta^{1} \:&
    \lambda_{3}{}^{1} \delta^{2} - \lambda_{3}{}^{2} \delta^{1}
    \end{pmatrix}_{ij} = \mu_{j}{}^{i} .
$$
The identities
\begin{equation}
  \vec{\mu} \vec{\lambda}^{t} = \begin{pmatrix} 0 & \delta^{3} & -\delta^{2} \\
    -\delta^{3} & 0 & \delta^{1} \\ \delta^{2} & -\delta^{1} & 0 \end{pmatrix},
    \:\:
  \vec{\mu} \vec{\mu}^{t} = \begin{pmatrix} (\delta^{2})^{2} + (\delta^{3})^{2}
    \:&  -\delta^{1} \delta^{2} \:& -\delta^{3} \delta^{1} \\ -\delta^{1}
    \delta^{2} \:& (\delta^{3})^{2} + (\delta^{1})^{2} \:& -\delta^{2}
    \delta^{3} \\ -\delta^{3} \delta^{1} \:& -\delta^{2} \delta^{3} \:&
    (\delta^{1})^{2} + (\delta^{2})^{2} \end{pmatrix} , \label{eq:cform:id}
\end{equation}
shall prove useful and, in addition, the inverse of \eref{eq:cform:paramtrans}
which reads
\begin{subequations}\begin{align}
  \vec{A} &= \vec{\lambda} \vec{\tilde{A}} \vec{\lambda}^{t} + 2\, \mathcal{S}
    (\vec{\lambda} \vec{\tilde{B}} \vec{\mu}^{t}) + \vec{\mu} \vec{\tilde{C}}
    \vec{\mu}^{t}, \label{eq:cform:invparamtransA} \\
  \vec{B} &= \vec{\lambda} \vec{\tilde{B}} \vec{\lambda}^{t} + \vec{\mu}
    \vec{\tilde{C}} \vec{\lambda}^{t}, \label{eq:cform:invparamtransB} \\
  \vec{C} &= \vec{\lambda} \vec{\tilde{C}} \vec{\lambda}^{t}.
    \label{eq:cform:invparamtransC}
\end{align}\label{eq:cform:invparamtrans}\end{subequations}

Our procedure for determining the transformation to canonical form for the
cases of translational and rotational CKTs is provided in
subsections~\ref{sec:cform:trans} and \ref{sec:cform:rot}, respectively. The
three asymmetric CKTs are each treated separately in
subsections~\ref{sec:cform:conical}--\ref{sec:cform:ellipsoidal}.

\subsection{The translational Killing tensors} \label{sec:cform:trans}

Let us consider first the Cartesian CKT. Since it is necessarily a constant
tensor, the transformation rules~\eref{eq:cform:paramtrans} reduce to
$\vec{\tilde{A}} = \vec{\lambda}^{t} \vec{A} \vec{\lambda}$, where
$\vec{\tilde{A}} = \mathrm{diag}(\tilde{a}_{1}, \tilde{a}_{2}, \tilde{a}_{3})$,
on account of \eref{eq:hj:w1}. Trivially, the $\tilde{a}_{i}$ are the
eigenvalues of $\vec{A}$, the columns of $\vec{\lambda}$ are the (normalized)
eigenvectors of $\vec{A}$ and the translation $\vec{\delta}$ is arbitrary.

Suppose now that the CKT is circular cylindrical, parabolic cylindrical or
elliptic-hyperbolic. We may assume without loss of generality that the tensor
has the form \eref{eq:hj:Ktrans}, since it must necessarily be of this
form in order to carry out the classification scheme in
section~\ref{sec:class}. Consequently, the rotation and translation are of the
form
$$
  \vec{\lambda} = \begin{pmatrix} \cos \phi \:& -\sin \phi \:& 0 \\ \sin \phi
    \:& \cos \phi \:& 0 \\ 0 \:& 0 \:& 1 \end{pmatrix}, \quad
  \vec{\delta} = \begin{pmatrix} \delta^{1} \\ \delta^{2} \\ 0 \end{pmatrix},
$$
and thus the problem reduces to finding $\phi$, $\delta^{1}$ and $\delta^{2}$.
The derivation of these parameters for each of the three translational
CKTs under consideration is provided in \cite{MST02} (see table~1, p~1432).
We now restate these results in our notation.

\begin{enumerate}
  \item Circular cylindrical case: The rotation angle $\phi$ is arbitrary and
    $$
      \delta^{1} = \frac{b_{23}}{c_{3}}, \quad
      \delta^{2} = -\frac{b_{13}}{c_{3}} .
    $$
  \item Parabolic cylindrical case: If $b_{23} \not= 0$, then
    $$
      \tan \phi = -\frac{b_{13}}{b_{23}},
    $$
    and $\phi = \frac{\pi}{2}$ for $b_{23} = 0$ (unique mod $\pi$). The
    components of the translation are
    $$
      \delta^{1} = \frac{b_{23}(a_{2} - a_{1}) + 2 \alpha_{3} b_{13}}
        {2(b_{13}{}^{2} + b_{23}{}^{2})}, \quad
      \delta^{2} = \frac{b_{13}(a_{2} - a_{1}) - 2 \alpha_{3} b_{23}}
        {2(b_{13}{}^{2} + b_{23}{}^{2})}.
    $$
  \item Elliptic-hyperbolic case: Let
    $$
      \sigma_{1} = b_{13}{}^{2} - b_{23}{}^{2} + c_{3} (a_{2} - a_{1}), \quad
      \sigma_{2} = \alpha_{3} c_{3} - b_{13} b_{23}, \quad
      \Delta = \sigma_{1}{}^{2} + 4 \sigma_{2}{}^{2}
    $$
    (and note that $\Delta$ is one of the fundamental
    invariants~\eref{eq:inv:KT2inv}). Then,
    \renewcommand{\arraystretch}{1.3}
    $$
      \tan\phi = \left\{ \begin{array}{ll}
        0, &\quad\text{if}\quad \text{$\sigma_{2} = 0$ and $\sigma_{1} < 0$,}\\
        \infty, &\quad\text{if}\quad \text{$\sigma_{2} = 0$ and
          $\sigma_{1} > 0$,} \\
        \frac{\sigma_{1} + \sqrt{\Delta}}{2 \sigma_{2}}, &\quad\text{if}
          \quad  \sigma_{2} \not= 0,  \end{array} \right.
    $$
    ($\phi$ unique mod $\pi$), and $\delta^{1}$ and $\delta^{2}$ are the
    same as in the circular cylindrical case. Moreover, the essential
    parameter $a$ satisfies
    $$
      a^{2} = \frac{\tilde{a}_{1}-\tilde{a}_{2}}{\tilde{c}_{3}} =
        \frac{\sqrt{\Delta}}{c_{3}{}^{2}}.
    $$\renewcommand{\arraystretch}{1.0}
\end{enumerate}

\subsection{The rotational Killing tensors} \label{sec:cform:rot}

By the same reasoning used in the previous subsection, we may assume without
loss of generality that the rotational Killing tensor has the form
\eref{eq:hj:Krot}. As the isometry group for this subspace of rotational webs
is the group of translations about the $z$-axis, we set $\lambda_{j}{}^{i} =
\delta_{j}{}^{i}$ and $\delta^{1} = \delta^{2} = 0$ in the transformation
rules~\eref{eq:cform:paramtrans}. The determination of $\delta^{3}$ and hence
the transformation to canonical form thus becomes a trivial calculation. It
follows that
$$
  \delta^{3} = \frac{b_{12}}{c_{2}}
$$
for the spherical, prolate spheroidal and oblate spheroidal cases and
$$
  \delta^{3} = \frac{a_{1} - a_{3}}{2 b_{12}}
$$
for the parabolic case. Finally, the essential parameter $a$ appearing in the
transformation from Cartesian to prolate (oblate) spheroidal coordinates
satisfies
$$
  a^{2} = \pm \frac{\tilde{a}_{3} - \tilde{a}_{1}}{\tilde{c}_{2}} =
    \pm \frac{\Delta_{2}}{\Delta_{1}{}^{2}},
$$
where $\Delta_{1}$ and $\Delta_{2}$ are the fundamental
$I(\mathbb{R})$-invariants~\eref{eq:inv:KR2inv} and the positive (negative)
signs correspond to the prolate (oblate) spheroidal tensor.

\subsection{The conical case} \label{sec:cform:conical}

The parameter matrices associated with the canonical form of the conical
Killing tensor specialize to
$$
  \vec{\tilde{A}} = \tilde{a}_{1} \vec{1}, \quad
  \vec{\tilde{B}} = \vec{0}, \quad
  \vec{\tilde{C}} = \mathrm{diag}(\tilde{c}_{1},\tilde{c}_{2},\tilde{c}_{3})
$$
(see \eref{eq:hj:w9}). From~\eref{eq:cform:paramtransC}, $\vec{\tilde{C}} =
\vec{\lambda}^{t} \vec{C} \vec{\lambda}$, hence, as in the Cartesian case,
the $\tilde{c}_{i}$ are the eigenvalues of $\vec{C}$ and the columns of
$\vec{\lambda}$ are the (normalized) eigenvectors of $\vec{C}$. It follows
that the essential parameters $b$ and $c$ satisfy
$$
  \frac{b^{2}}{c^{2}} = \frac{\tilde{c}_{2} - \tilde{c}_{1}}{\tilde{c}_{3} -
    \tilde{c}_{1}},
$$
thus, in order to satisfy the condition $b^{2} < c^{2}$, we can order the
eigenvalues such that $\tilde{c}_{1} < \tilde{c}_{2} < \tilde{c}_{3}$.
Finally, substituting \eref{eq:cform:paramtransC} into
\eref{eq:cform:invparamtransB} leads to $\vec{B} = \vec{\mu} \vec{\lambda}^{t}
\vec{C}$, which can easily be solved for $\vec{\delta}$ noting the
identity~\eref{eq:cform:id}.

\subsection{The paraboloidal case} \label{sec:cform:paraboloidal}

The parameter matrices associated with the canonical form of the paraboloidal
Killing tensor specialize to
\begin{equation}
  \vec{\tilde{A}} = \mathrm{diag}(\tilde{a}_{1},\tilde{a}_{2},\tilde{a}_{3}),
  \quad \vec{\tilde{B}} = \begin{pmatrix} 0 & \tilde{b}_{12} & 0 \\
  \tilde{b}_{21} & 0 & 0 \\ 0 & 0 & 0 \end{pmatrix}, \quad
  \vec{\tilde{C}} = \mathrm{diag}(0,0,\tilde{c}_{3}),
    \label{eq:cform:paraboloidal1}
\end{equation}
subject to the constraint
\begin{equation}
  \tilde{b}_{12}[\tilde{b}_{12} \tilde{b}_{21} + \tilde{c}_{3}
    (\tilde{a}_{2} - \tilde{a}_{3})] + \tilde{b}_{21}[\tilde{b}_{12}
    \tilde{b}_{21} + \tilde{c}_{3}(\tilde{a}_{1} - \tilde{a}_{3})] = 0
    \label{eq:cform:paraboloidal2}
\end{equation}
(see \eref{eq:hj:w10}). It follows that the essential constants $b$ and $c$
satisfy
\begin{equation}\renewcommand{\arraystretch}{1.3}
  \left\{ \begin{array}{lll}
    c = \frac{\tilde{a}_{2}-\tilde{a}_{3}}{2\tilde{b}_{12}}, &\quad
      c - b = \frac{\tilde{a}_{2}-\tilde{a}_{1}}{2\tilde{b}_{12}}, &\quad
      \text{if}\quad \tilde{c}_{3} = 0, \\
    b = \frac{\tilde{a}_{1}-\tilde{a}_{2}}{2\tilde{b}_{12}}, &\quad
      c - b = \frac{\tilde{b}_{12}}{2\tilde{c}_{3}}, &\quad
      \text{if}\quad \tilde{c}_{3} \not= 0, \: \tilde{b}_{21} = 0, \\
    c = \frac{\tilde{a}_{1}-\tilde{a}_{2}}{2\tilde{b}_{21}}, &\quad
      c - b = \frac{\tilde{b}_{21}}{2\tilde{c}_{3}}, &\quad
      \text{if}\quad \tilde{c}_{3} \not= 0, \: \tilde{b}_{12} = 0, \\
    b = \frac{\tilde{a}_{1}-\tilde{a}_{3}}{2\tilde{b}_{12}}, &\quad
      c - b = \frac{\tilde{b}_{12}+\tilde{b}_{21}}{2\tilde{c}_{3}}, &\quad
      \text{if}\quad \tilde{c}_{3} \not= 0, \: \tilde{b}_{12} \not= 0, \:
      \tilde{b}_{21} \not= 0,
  \end{array} \right. \label{eq:cform:paraboloidal3}
\renewcommand{\arraystretch}{1.0}\end{equation}
together with the condition $b > c$. From~\eref{eq:cform:paramtransC},
$\vec{\tilde{C}} = \vec{\lambda}^{t} \vec{C} \vec{\lambda}$, hence it follows
from \eref{eq:cform:paraboloidal1} that $\vec{C}$ necessarily has a zero
eigenvalue of multiplicity two and one other eigenvalue $\tilde{c}_{3}$.
We now consider the two cases $\tilde{c}_{3} = 0$ and $\tilde{c}_{3} \not= 0$
separately.

If $\tilde{c}_{3} = 0$, then it follows from \eref{eq:cform:paraboloidal2} that
$\tilde{b}_{21} = -\tilde{b}_{12}$. Moreover, \eref{eq:cform:paramtransC} is
trivially satisfied and \eref{eq:cform:paramtransB} reduces to $\vec{\tilde{B}}
= \vec{\lambda}^{t} \vec{B} \vec{\lambda}$. This implies that
$\vec{\tilde{B}}^{2} = \vec{\lambda}^{t} \vec{B}^{2} \vec{\lambda}$, where
$\vec{\tilde{B}}^{2} = \mathrm{diag}(-\tilde{b}_{12}{}^{2},
-\tilde{b}_{12}{}^{2}, 0)$. Therefore, the negative eigenvalue of $\vec{B}^{2}$
determines $\tilde{b}_{12}$; we can take its sign to be positive without loss
of generality. The normalized eigenvectors of $\vec{B}^{2}$ determine
$\vec{\lambda}$ up to a rotation in the eigenspace associated with the negative
eigenvalue which fixes $\vec{\lambda}$ up to a parameter $\psi$. Finally, it
follows that \eref{eq:cform:invparamtransA} reduces to $\vec{A} = 2\,
\mathcal{S}(\vec{B} \vec{\lambda} \vec{\mu}^{t}) + \vec{\lambda}
\vec{\tilde{A}} \vec{\lambda}^{t}$. This equation can be solved for the
$\tilde{a}_{i}$, $\delta^{i}$, and $\psi$ which, in general, yields multiple
solutions; a particular solution satisfying the condition $b > c$
(see~\eref{eq:cform:paraboloidal3}) can be selected.

Finally, if $\tilde{c}_{3} \not= 0$, then the eigenproblem
\eref{eq:cform:paramtransC} uniquely determines $\tilde{c}_{3}$ and
$\vec{\lambda}$ up to a parameter $\psi$.
Equations~\eref{eq:cform:invparamtransA} and \eref{eq:cform:invparamtransB}
can then be solved for $\tilde{b}_{12}$, $\tilde{b}_{21}$, $\tilde{a}_{i}$,
$\delta^{i}$, and $\psi$, in conjunction with the condition $b > c$.

\subsection{The ellipsoidal case} \label{sec:cform:ellipsoidal}

The parameter matrices associated with the canonical form of the ellipsoidal
Killing tensor specialize to
\begin{equation}
  \vec{\tilde{A}} = \mathrm{diag}(\tilde{a}_{1},\tilde{a}_{2},\tilde{a}_{3}),
  \quad \vec{\tilde{B}} = \mathbf{0}, \quad
  \vec{\tilde{C}} = \mathrm{diag}(\tilde{c}_{1},\tilde{c}_{2},\tilde{c}_{3}),
    \label{eq:cform:ellipsoidal1}
\end{equation}
subject to the constraint
\begin{equation}
  (\tilde{a}_{1} - \tilde{a}_{2}) \tilde{c}_{1} \tilde{c}_{2} +
  (\tilde{a}_{2} - \tilde{a}_{3}) \tilde{c}_{2} \tilde{c}_{3} +
  (\tilde{a}_{3} - \tilde{a}_{1}) \tilde{c}_{3} \tilde{c}_{1} = 0
    \label{eq:cform:ellipsoidal2}
\end{equation}
(see \eref{eq:hj:w11}). It follows that the essential constants $a$, $b$
and $c$ satisfy
\begin{equation}\renewcommand{\arraystretch}{1.3}
  \left\{ \begin{array}{lll}
    a - b = \frac{\tilde{a}_{1}-\tilde{a}_{2}}{\tilde{c}_{3}}, &\quad
      c - a = \frac{\tilde{a}_{3}-\tilde{a}_{1}}{\tilde{c}_{2}}, &\quad
      \text{if}\quad \tilde{c}_{2} \not= 0, \: \tilde{c}_{3} \not= 0, \\
    a - b = \frac{\tilde{a}_{1}-\tilde{a}_{2}}{\tilde{c}_{3}}, &\quad
      c - b = \frac{\tilde{a}_{1}-\tilde{a}_{2}}{\tilde{c}_{1}}, &\quad
      \text{if}\quad \tilde{c}_{2} = 0, \: \tilde{c}_{3} \not= 0, \\
    c - a = \frac{\tilde{a}_{3}-\tilde{a}_{1}}{\tilde{c}_{2}}, &\quad
      c - b = \frac{\tilde{a}_{3}-\tilde{a}_{1}}{\tilde{c}_{1}}, &\quad
      \text{if}\quad \tilde{c}_{2} \not= 0, \: \tilde{c}_{3} = 0,
  \end{array} \right. \label{eq_cform_ellipsoidal3}
\end{equation}
together with the condition $a > b > c$. As in the conical case,
\eref{eq:cform:paramtransC} implies that the $\tilde{c}_{i}$ are the
eigenvalues of $\vec{C}$ and the columns of $\vec{\lambda}$ are the
(normalized) eigenvectors of $\vec{C}$. There are two cases to consider: (1)
the $\tilde{c}_{i}$ are all distinct and (2) the $\tilde{c}_{i}$ are all
equal\footnote{The case of only two equal $\tilde{c}_{i}$ is impossible,
for such a Killing tensor would characterize either an elliptic-hyperbolic,
prolate spheroidal or oblate spheroidal web.}.

If the $\tilde{c}_{i}$ are all distinct, then the matrix $\vec{\lambda}$ is
uniquely determined (up to the ordering of the columns) and $\vec{\delta}$ can
be computed as in the conical case. Upon obtaining $\vec{\tilde{A}}$ and
$\vec{\tilde{C}}$, the condition $a > b > c$ should be verified and, if
necessary, the eigenvalues $\tilde{c}_{i}$ may need to be reordered
accordingly.

If all of the $\tilde{c}_{i}$ are equal, then \eref{eq:cform:paramtransC} is
trivially satisfied. Equation~\eref{eq:cform:invparamtransB} reduces to
$\vec{B} = \tilde{c}_{1} \vec{\mu} \vec{\lambda}^{t}$ which can be solved for
$\vec{\delta}$ using the identity~\eref{eq:cform:id}. Finally,
\eref{eq:cform:invparamtransA} simplifies to $\vec{A} - \tilde{c}_{1} \vec{\mu}
\vec{\mu}^{t} = \vec{\lambda} \vec{\tilde{A}} \vec{\lambda}^{t}$. This
eigenproblem can be solved to obtain $\vec{\lambda}$.


\section{Main algorithm} \label{sec:alg}

The above-presented considerations lead to a systematic and computationally
efficient method of determining separable coordinates for the natural
Hamiltonian~\eref{eq:hj:natH}. We emphasize that our algorithm is
\textit{purely algebraic}, and hence is well suited for implementation in a
symbolic computer algebra system. Indeed, as we mentioned in
section~\ref{sec:intro}, the algorithm has been fully implemented into Maple
through the \texttt{KillingTensor} package. We now summarize the three main
steps of the algorithm.

\textit{(1) Impose the compatibility condition.} Using the given potential $V$
in terms of Cartesian coordinates $x^{i}$ and a generic Killing tensor
$\vec{K}$ of the form \eref{eq:ITK2E3:Ktens_compts2}, impose the compatibility
condition~\eref{eq:hj:dKdV} to obtain the equivalent conditions on the Killing
tensor parameters~\eref{eq:ITK2E3:KTparams}. Computationally, this step amounts
to solving a system of linear equations in the
parameters~\eref{eq:ITK2E3:KTparams}.

\textit{(2) Extract the orthogonal coordinate webs.} Decompose the general
solution obtained in step~(1) into the form
\begin{equation}
  \vec{K} = \ell_{0} \vec{g} + \ell_{1} \vec{K}_{1} + \cdots + \ell_{n}
    \vec{K}_{n}, \label{eq:alg:decomp}
\end{equation}
where $\ell_{i}$, $i = 1, \ldots, n$ are arbitrary constants, $\vec{g}$ is the
metric tensor and $\{\vec{K}_{1}, \ldots, \vec{K}_{n}\}$ is a linearly
independent set of Killing tensors, noting that $n \leq 19$ since
$\mathrm{dim}\, \mathcal{K}^{2}(\Eth) = 20$. By theorem~4.1, each $\vec{K}_{i}$
must necessarily have normal eigenvectors and distinct eigenvalues if it is to
characterize separation in one of the eleven separable coordinate systems. The
former can be verified using the TSN conditions~\eref{eq:hj:tsn} while the
latter can be verified efficiently by computing the discriminant of the
characteristic polynomial of $\vec{K}_{i}$ and checking that it does not vanish
identically. Finally, relabel the $\vec{K}_{i}$ so that $\vec{K}_{1}, \ldots,
\vec{K}_{m}$, $m \leq n$, are CKTs.

\textit{(3) Classify each Killing tensor and transform to canonical form.} For
each $\vec{K}_{i}$, $i=1, \ldots, m$, in step~(2), classify $\vec{K}_{i}$ using
the scheme in section~\ref{sec:class}. Finally, using the techniques described
in section~\ref{sec:cform}, determine the
transformation~\eref{eq:cform:trans_x} which brings $\vec{K}_{i}$ to its
appropriate canonical form. The transformation to separable coordinates can be
carried out using equation~\eref{eq:cform:trans_sep}.\medskip

\noindent \textit{Remark.} Because of the non-linearity of the integrable
eigenvector and distinct eigenvalue conditions as well as the non-linearity of
the fundamental invariants, certain linear combinations of the $\vec{K}_{i}$,
$i=1, \ldots, n$ in step~(2) may produce Killing tensors which characterize
separability in coordinate systems not characterized by the $\vec{K}_{i}$,
$i=1, \ldots, m$. In fact, it may be possible to construct such a linear
combination where the individual Killing tensors of the combination fail to
have normal eigenvectors or distinct eigenvalues. This will be illustrated by
the example in the next section.


\section{Application: The Calogero-Moser system} \label{sec:cm}

We apply the algorithm of section~\ref{sec:alg} to the three-body inverse
square Calogero-Moser system with equal masses. It is defined by the
natural Hamiltonian~\eref{eq:hj:natH} with potential
\begin{equation}
  V = \frac{1}{(x-y)^{2}} + \frac{1}{(y-z)^{2}} + \frac{1}{(z-x)^{2}}.
    \label{eq:cm:potential}
\end{equation}
Solving the compatibility condition~\eref{eq:hj:dKdV} with the
potential~\eref{eq:cm:potential} yields
\begin{equation}
  \vec{K} = a_{1} \vec{g} + \alpha_{1} \vec{K}_{1} + b_{32} \vec{K}_{2}
    + c_{3} \vec{K}_{3} + \gamma_{3} \vec{K}_{4}, \label{eq:cm:Kcompat1}
\end{equation}
where
\begin{equation}\begin{split}
  K^{ij}_{1} = \begin{pmatrix} 0 & 1 & 1 \\ 1 & 0 & 1 \\ 1 & 1 & 0
    \end{pmatrix}, \quad
  K^{ij}_{4} = \begin{pmatrix} -2yz \:& (x+y-z)z \:& (z+x-y)y \\ (x+y-z)z
    \:& -2zx \:& (z+y-x)x \\ (z+x-y)y \:& (z+y-x)x \:& -2xy \end{pmatrix}, \\
  K^{ij}_{2} = \begin{pmatrix} 2 y + 2 z \:& -x - y \:& -z - x \\
    -x - y \:& 2 z + 2 x \:& -y - z \\ -z - x \:& -y - z \:&
    2 x + 2 y \end{pmatrix}, \quad
  K^{ij}_{3} = \begin{pmatrix} y^{2} + z^{2} & -xy & -zx \\ -xy & z^{2}
    + x^{2} & -yz \\ -zx & -yz & x^{2} + y^{2} \end{pmatrix} .
\end{split}\label{eq:cm:Kcompat2}\end{equation}
Using~\eref{eq:hj:tsn}, we find that $\vec{K}$ has normal eigenvectors for all
$a_{1}$, $\alpha_{3}$, $b_{32}$, $c_{3}$ and $\gamma_{3}$. However, it follows
that only $\vec{K}_{2}$ and $\vec{K}_{4}$ have distinct eigenvalues. It is also
useful to note that $\vec{K}$ in \eref{eq:cm:Kcompat1} admits a Killing vector
\begin{equation}
  \vec{V} = (y-z) \vec{X}_{1} + (z-x) \vec{X}_{2} + (x-y) \vec{X}_{3} .
    \label{eq:cm:Kvec}
\end{equation}
Using the Killing vector classification scheme in
section~\eref{sec:class:K1R3}, it follows from table~\eref{sec:class:K1R3} that
$\vec{V}$ is rotational and the transformation
\begin{equation}
  x^{i} = \lambda_{j}{}^{i} \tilde{x}^{j} + \delta^{i}, \quad
  \lambda_{j}{}^{i} = \frac{1}{\sqrt{6}} \begin{pmatrix} 2 & 0 & \sqrt{2} \\
    -1 & \sqrt{3} & \sqrt{2} \\ -1 & -\sqrt{3} & \sqrt{2} \end{pmatrix}_{ij},
    \quad \delta^{i} = 0, \label{eq:cm:trans}
\end{equation}
brings \eref{eq:cm:Kvec} to the canonical form \eref{eq:class:K1R3c}. Let us
now apply step~(3) of the algorithm in section~\ref{sec:alg} to each of the
CKTs in \eref{eq:cm:Kcompat2}.

Applying the transformation~\eref{eq:cm:trans} to $K^{ij}_{2}$ yields
$$
  \tilde{K}^{ij}_{2} = \sqrt{3} \begin{pmatrix} 2 \tilde{z} & 0
    & -\tilde{x} \\ 0 & 2 \tilde{z} & -\tilde{y} \\ -\tilde{x} & -\tilde{y}
    & 0 \end{pmatrix}.
$$
It follows that $\vec{\tilde{K}}_{2} \in \mathcal{K}_{R}^{2}(\Eth)$ with
respect to the transformed Cartesian coordinates $\tilde{x}^{i}$, and thus, by
table~\ref{table:class:KR2E3}, we see that $\vec{K}_{2}$ characterizes a
parabolic web. Finally, we observe from \eref{eq:hj:w8} that
$\tilde{K}^{ij}_{2}$ is already in canonical form. Therefore, it follows from
\eref{eq:cm:trans} together with \eref{eq:hj:sc8} that the transformation to
separable parabolic coordinates $(\mu,\nu,\psi)$ is given by
\begin{align*}
  x &= \tfrac{\sqrt{2}}{\sqrt{3}} \mu\nu\cos\psi + \tfrac{1}{2\sqrt{3}}
    (\mu^{2} - \nu^{2}), \\
  y &= -\tfrac{1}{\sqrt{6}} \mu\nu\cos\psi + \tfrac{1}{\sqrt{2}} \mu\nu\sin\psi
    + \tfrac{1}{2\sqrt{3}}(\mu^{2} - \nu^{2}), \\
  z &= -\tfrac{1}{\sqrt{6}} \mu\nu\cos\psi - \tfrac{1}{\sqrt{2}} \mu\nu\sin\psi
    + \tfrac{1}{2\sqrt{3}}(\mu^{2} - \nu^{2}).
\end{align*}

Proceeding similarly for $\vec{K}_{4}$, it follows that it enjoys the form
$$
  \tilde{K}^{ij}_{4} = \begin{pmatrix} 2 \tilde{y}^{2} - \tilde{z}^{2} &
    -2 \tilde{x} \tilde{y} & \tilde{z} \tilde{x} \\ -2 \tilde{x} \tilde{y} &
    2 \tilde{x}^{2} - \tilde{z}^{2} & \tilde{y} \tilde{z} \\ \tilde{z}
    \tilde{x} & \tilde{y} \tilde{z} & -\tilde{x}^{2} - \tilde{y}^{2}
    \end{pmatrix} ,
$$
under the transformation~\eref{eq:cm:trans}. We conclude from
table~\ref{table:class:KR2E3} that $\tilde{K}^{ij}_{4}$ characterizes a
spherical web and is in canonical form upon comparison with \eref{eq:hj:w5}.
Thus, the transformation to separable spherical coordinates is given by
\eref{eq:cm:trans} together with \eref{eq:hj:sc5}.

As mentioned in the remark at the end of section~\ref{sec:alg}, we can take
various linear combinations of $\vec{K}_{1}, \ldots, \vec{K}_{4}$ in an
attempt to find additional separable coordinate systems. For example, let
$\vec{K}_{5,6} = \vec{K}_{3} \pm \vec{K}_{1}$. It follows that $\vec{K}_{5,6}$
both have distinct eigenvalues (even though $\vec{K}_{1}$ does not). Applying
the transformation~\eref{eq:cm:trans} to $\vec{K}_{5,6}$ yields
$$
  \tilde{K}^{ij}_{5,6} = \begin{pmatrix} \mp 1 + \tilde{y}^{2} + \tilde{z}^{2}
    & -\tilde{x} \tilde{y} & -\tilde{z} \tilde{x} \\ -\tilde{x} \tilde{y}
    & \mp 1 + \tilde{z}^{2} + \tilde{x}^{2} & -\tilde{y} \tilde{z} \\
    -\tilde{z} \tilde{x} & -\tilde{z} \tilde{y} & \pm 2 + \tilde{x}^{2}
    + \tilde{y}^{2} \end{pmatrix} .
$$
>From table~\ref{table:class:KR2E3} we conclude that $\vec{K}_{5}$ and
$\vec{K}_{6}$ characterize the prolate spheroidal and oblate spheroidal
webs, respectively. Moreover, $\tilde{K}^{ij}_{5,6}$ is in canonical
form (compare with \eref{eq:hj:w7} and \eref{eq:hj:w8}) and the essential
parameter appearing in the transformation from Cartesian to prolate (or oblate)
spheroidal coordinates is $a = \sqrt{3}$ for both cases (see
section~\ref{sec:cform:rot}).

Finally, it follows that $\vec{K}_{7} = \vec{K}_{1} + \vec{K}_{3} +
\vec{K}_{4}$ has distinct eigenvalues and admits a translational Killing vector
\begin{equation}
  \vec{V} = \vec{X}_{1} + \vec{X}_{2} + \vec{X}_{3} . \label{eq:cm:Kvec_trans}
\end{equation}
It follows that the transformation~\eref{eq:cm:trans} brings
\eref{eq:cm:Kvec_trans} to the canonical form \eref{eq:class:K1R3b}. Applying
this transformation to $K^{ij}_{7}$ yields
$$
  \tilde{K}^{ij}_{7} = \begin{pmatrix} -1 + 3 \tilde{y}^{2} & -3 \tilde{x}
    \tilde{y} & 0 \\ -3 \tilde{x} \tilde{y} & -1 + 3 \tilde{x}^{2} & 0 \\
    0 & 0 & 2 \end{pmatrix},
$$
which characterizes the circular cylindrical web.

To summarize, we have shown using the algorithm of section~\ref{sec:alg} that
the Calogero-Moser system with potential~\eref{eq:cm:potential} separates in
five orthogonally separable coordinate systems, namely, circular cylindrical,
spherical, prolate spheroidal, oblate spheroidal and parabolic. This result is
consistent with that found in \cite{BCR00} and \cite{C01}.

We conclude by making two remarks. Firstly, our analysis is \textit{exhaustive}
in the sense that we have found \textit{all} possible orthogonally separable
coordinate systems for which the Calogero-Moser system separates, since it
follows that the general Killing tensor~\eref{eq:cm:Kcompat1} admits the
rotational Killing vector~\eref{eq:cm:Kvec} and hence can only characterize a
rotational web. Secondly, the same conclusions from this section hold for a
weighted Calogero-Moser system with unequal masses. More precisely, the
Hamiltonian system~\eref{eq:hj:natH} with potential
$$
  V = \frac{g_{1}}{(m_{1} x - m_{2} y)^{2}} + \frac{g_{2}}{(m_{2} y
    - m_{3} z)^{2}} + \frac{g_{3}}{(m_{3} z - m_{1} x)^{2}}
$$
separates in the five aforementioned coordinate systems, where $g_{i}$ and
$m_{i}$ are constants and $m_{i} > 0$. Moreover, in all five cases, the
transformation to separable coordinates is given by
$$
  x^{i} = \lambda_{j}{}^{i} T^{j}(u^{k}) + \delta^{i}, \quad
  \lambda_{j}{}^{i} = \begin{pmatrix} m_{1} M N^{-1} \:& 0 \:& m_{2} m_{3} M
    \\ -m_{2} m_{3}{}^{2} M N \:& m_{2} N \:& m_{3} m_{1} M \\ -m_{3}
    m_{2}{}^{2} M N \:& -m_{3} N \:& m_{1} m_{2} M \end{pmatrix}_{ij}, \quad
  \delta^{i} = 0,
$$
where
$$
  M = (m_{1}{}^{2} m_{2}{}^{2} + m_{2}{}^{2} m_{3}{}^{2} + m_{3}{}^{2}
    m_{1}{}^{2})^{-1/2}, \quad N = (m_{2}{}^{2} + m_{3}{}^{2})^{-1/2} .
$$
The authors believe that this result is new.


\section{Conclusions} \label{sec:conc}

In this paper we solve a non-trivial problem of the geometry of orthogonal
coordinate webs, namely the classification of the eleven orthogonal coordinate
webs in $\Eth$ in terms of the invariants of the corresponding vector spaces of
Killing two-tensors and vectors. Notably, the original solution presented here
fits well the approach to geometry of Felix Klein presented in his Erlangen
Program. Moreover, the results are successfully applied to the integrability
problem of the Hamiltonian systems defined in $\Eth$. From this viewpoint, the
well-known Calogero-Moser super-separable Hamiltonian system has been
integrated within the framework of the (orthogonal) Hamilton-Jacobi theory of
separation of variables. The other three-dimensional pseudo-Riemannian flat
space that is amendable to the methods developed in this work is Minkowski
space $\mathbb{E}^{2,1}$. The work in this direction is underway.


\section*{Acknowledgements}

The authors wish to thank the Department of Mathematics, University of Turin
for hospitality during which part of this paper was written. They wish to
express their appreciation for helpful discussions with Sergio Benenti, Claudia
Chanu, Robin Deeley, Lorenzo Fatibene, Giovanni Rastelli and Dennis The. The
research was supported in part by Natural Sciences and Engineering Research
Council of Canada (NSERC) Discovery Grants (RGM, RGS), an NSERC Postgraduate
Scholarship (JTH) and by a Senior Visiting Professorship of the Gruppo
Nazionale di Fisica Matematica dell'Italia (RGM).


\end{document}